\documentclass[aps,twocolumn,pra,superscriptaddress]{revtex4-2}

\usepackage{amsmath}
\usepackage{graphicx}
\usepackage{stackengine}
\usepackage{amsfonts}
\usepackage{braket}
\usepackage{mathtools}
\usepackage[toc]{appendix}
\usepackage[colorlinks=true,linkcolor=blue,urlcolor=blue,citecolor=blue,pdfusetitle]{hyperref}
\usepackage[all]{hypcap} 
\usepackage{ulem,tabularx,booktabs}
\usepackage{comment,palatino}
\newcolumntype{Y}{>{\centering\arraybackslash}X}

\newcommand{\eq}[1]{\text{Eq.}~\eqref{#1}}
\newcommand{\tab}[1]{\text{Tab.}~\ref{#1}}
\newcommand{\fig}[1]{\text{Fig.}~\ref{#1}}

\newcommand{\subf}[2]{\text{Fig.}~\ref{#1}\textcolor{blue}{#2}}

\newcommand{\sect}[1]{\text{Sec.}~\ref{#1}}

\newcommand{\ie}{\textit{i}.\textit{e}.}
\newcommand{\eg}{\textit{e}.\textit{g}.}

\newcommand{\tr}[1]{\text{Tr}\left\{#1\right\}}

\newcommand{\mC}{\mathcal{C}}
\newcommand{\mD}{\mathcal{D}}
\newcommand{\mJ}{\mathcal{J}}
\newcommand{\mL}{\mathcal{L}}
\newcommand{\mQ}{\mathcal{Q}}
\newcommand{\mQF}{\mathcal{G}}
\newcommand{\mU}{\mathcal{U}}
\newcommand{\kB}{k_\text{\tiny B}}
\newcommand{\D}{\operatorname{d\!}}
\newcommand{\abs}[1]{\lvert #1 \rvert}
\DeclarePairedDelimiter{\norm}{\lVert}{\rVert}

\graphicspath{{Figures/}}

\begin{document}
\title{
Quantifying protocol efficiency: a thermodynamic figure of merit for classical and quantum state-transfer protocols}
\author{Qiongyuan Wu}
\affiliation{Centre for Quantum Materials and Technologies, School of Mathematics and Physics, Queen’s University Belfast, BT7 1NN Belfast, UK}
\author{Mario A. Ciampini}
\affiliation{University of Vienna, Faculty of Physics, Vienna Center for Quantum Science and Technology (VCQ), A-1090 Vienna, Austria}
\author{Mauro Paternostro}
\affiliation{Centre for Quantum Materials and Technologies, School of Mathematics and Physics, Queen’s University Belfast, BT7 1NN Belfast, UK}
\author{Matteo Carlesso}
\thanks{Email: m.carlesso@qub.ac.uk}
\affiliation{Centre for Quantum Materials and Technologies, School of Mathematics and Physics, Queen’s University Belfast, BT7 1NN Belfast, UK}

\date{\today}

\begin{abstract}
Manipulating quantum systems undergoing non-Gaussian dynamics in a fast and accurate manner is  becoming fundamental to many quantum applications. Here, we focus on classical and quantum protocols transferring a state across a double-well potential.
The classical protocols are achieved by deforming the potential, while the quantum ones are assisted by a counter-diabatic driving. We show that quantum protocols perform more quickly and accurately. Finally, we design a figure of merit for the performance of the transfer protocols -- namely, the \textit{protocol grading} -- that depends only on fundamental physical quantities, and which accounts for the quantum speed limit, the fidelity and the thermodynamic of the process. We test the protocol grading with classical and  quantum protocols, and show that quantum protocols have higher protocol grading than the classical ones.
\end{abstract}

\maketitle

\section{Introduction}\label{sec:introduction}

Advances in the manipulation of Gaussian states and dynamics~\cite{Wang_2007,Weedbrook_2012, Yokoyama_2013, Roslund_2014, Asavanant_2019} have enabled 
experimental achievement in quantum sensing \cite{Giovannetti_2004,Pirandola_2018}, communication and computation \cite{Gisin_2007,Zhong_2020,Chen_2021}. 
However, a framework based only on Gaussian states is inadequate for universal quantum computation~\cite{Holevo_1999,Wolf_2006}, for instance, 
with the availability of suitable non-Gaussian resources being a necessary ingredient to unlock the potential of continuous-variable quantum information processing~\cite{Lloyd_1999,Gottesman_2001,Lee_2011, Lvovsky_2020,Walschaers_2021}.
Recent improvements have demonstrated the viability of generation and manipulation of non-Gaussian systems in platforms such as non-linear optics, ultracold atoms \cite{Amico_2021}, (opto- and electro-)mechanical systems \cite{Millen_2020,Xu_2022}, and super-conducting systems \cite{Kjaergaard_2020}. A simple, nearly platform-agnostic paradigm for the engineering of non-Gaussian states is embodied by the {\it double-well potential} ~\cite{Novaes_2003,Kierig_2008}, whose richness and effectiveness have been proven key in applications of quantum information processing and quantum thermodynamics \cite{Gavrilov_2016,Campbell_2016}.

Quantum control can be successfully deployed in the quest for the generation~\cite{Shi_2019,Davis_2021,Ma_2021} and manipulation~\cite{Giannelli_2022,Koch_2022} of non-Gaussian systems. Well-known techniques, from feedback control \cite{PhysRevLett.117.163601, Rossi:2018ve,Delic:2020us} to optimal control \cite{Boscain_2021} and shortcut-to-adiabacity (STA) \cite{del_Campo_2019,Guery-Odelin_2019} are generally designed to optimize different aspects of a  quantum process, while the possibility to achieve enhanced performances -- above and beyond those characterizing their classical counterparts \cite{Giovannetti_2011,Han-Sen_2020,Hammam_2021,Ji_2022} -- through the combination of multiple techniques has been investigated~\cite{Stefanatos_2013,Martikyan_2020,Zhang_2021}. Among them, STA protocols aim at minimising the leakage into high-excitation subspaces that would be inevitably entailed by the fast dynamics of a quantum system, thus mimicking an {\it adiabatic} process that would otherwise be unachievable. A relevant form of STA protocols is counter-diabatic driving (CD) \cite{Berry_2009,del_Campo_2013}, which has acquired popularity owing to its simple structure that makes it ready implementable in problems of system translation \cite{An_2016,Sels_2017}, state engineering \cite{Abah_2020,Simon_2020}, and open systems dynamics \cite{Alipour_2020,Yin_2022}. 

The development of effective control techniques has not been accompanied by a concomitant effort aimed at the identification of comprehensive quantifiers to measure the advantages and costs of implementing quantum control. Yet, the availability of such a figure of merit -- which should allow the characterization of the quality of a protocol -- 
would allow the comparison of performances of different processes implemented with different approaches. 

Motivated by the need to provide a physically motivated figure of merit that is able to capture the facets of a quantum control protocol and inform against relevant performance indicators, here we 
put forward a quantifier, which we dub \textit{protocol grading parameter}, built around fundamental quantities such as speed of performance of a protocol, fidelity of implementation, and entropic cost. Our proposed tool is able to holistically assess the implications of embedding quantum control approaches into a given dynamical process, thus informing the selection of the best protocol to apply to a given problem among different ones that might be available. We benchmark our proposal by applying a CD scheme to a quantum system trapped in a double-well potential and addressing  the performance of (accelerated) quantum control-enhanced state transfer. The choice of our paradigm problem is relevant from a number of relevant viewpoints. First, by reinterpreting the process of transferring a quantum mechanical system across a double-well as the backbone of a {\it logical resetting} mechanism, our analysis can provide a characterisation of any logical operation that does not rely explicitly on Landauer-like arguments but focuses on an inherently dynamical approach. Second, by assessing speed, reliability and energetic cost of dynamical switching, our figure of merit and study will be instrumental to the furthering of the current effort dedicated to the characterization of quantum memories \cite{Lvovsky_2009,Hedges_2010,Heshami_2016}, providing key information for their experimental implementation.

The remainder of this paper is structured as follows.
In \sect{sec:quantifier}, we design the protocol grading to quantify the performances of such protocols, and discuss the fundamental quantities used as its building blocks.
In \sect{sec:protocols}, we define the task of transferring the state of a quantum system across a double-well potential, and introduce four protocols for its implementation. 
In \sect{sec:numericalsimulation}, we simulate the  protocols using methods developed in Ref.~\cite{Qiongyuan_2022}, while we discuss their performance in \sect{sec:quantification}.  Finally, in \sect{sec:conclusions} we draw our conclusions.

\section{Protocol grading}\label{sec:quantifier}

The main aim of this work is to grade the protocols through the newly introduced \textit{protocol grading parameter}, which we will indicate with $\mQF$, based on the following idea. If the state transfer protocol: 1) performs quickly, 2) consumes a small amount of energy, 3) produces faithful results, and 4) has a low degree of irreversibility, it would be considered as successful. Correspondingly, $\mQF$ would achieve its largest possible value (in the following, we shall normalize our figure of merit so that  $\mQF\in[0,1]$). Conversely, if the above performance indicators are not met, the protocol will be considered to perform poorly, and we associate it to a low value of $\mQF$ (close to $0$). 
Our analysis thus explicitly takes into consideration the quantum speed limit $g_\text{\tiny S}$, experimental quality $g_\text{\tiny Q}$, and thermodynamic cost $g_\text{\tiny T}$ of the protocol. It is thus natural to construct the quantity
\begin{equation}\label{equ:quantifier}
  \mQF=g_\text{\tiny S} g_\text{\tiny Q} g_\text{\tiny T}.
\end{equation}
The form of $g_\text{\tiny S}$, $g_\text{\tiny Q}$ and $g_\text{\tiny T}$ are summarised in \tab{tab:quantifier}, and they are explicitly discussed below. We can anticipate that they depend only on universal fundamental quantities, and thus $\mQF$ can be employed to quantify the quality of the transfer protocol independently from the scheme and the platform where is performed.
\begin{table}[t]
  \centering
  \begin{tabular}{c|c}
     \rule{0mm}{3mm}
    Quantity & Definition \\
    \hline
    \hline
    \rule{0mm}{5mm}
    $g_\text{\tiny S}$ & $~\max \left\{ 0,~ 1-0.1\times\log_{10}\left(\frac{\tau}{\tau_\text{\tiny QSL}}\right) \right\}~$ \\[1.5ex]
    \hline
    \rule{0mm}{5mm}
    $g_\text{\tiny Q}$ & $~F_\text{exp}(\rho_\text{f},\rho_\text{\tiny TG})~$ \\[1.5ex]
    \hline
    \rule{0mm}{5mm}
    $g_\text{\tiny T}$ & $~\exp{(-\Sigma_\text{ir})}~$ \\[1.5ex]
    \hline
  \end{tabular}
  \caption{\label{tab:quantifier} Definitions of the terms entering the protocol grading $\mQF$ that is defined in \eq{equ:quantifier}: $g_\text{\tiny S}$ describes the energetic cost, $g_\text{\tiny Q}$ expresses the experimental quality, while $g_\text{\tiny T}$ accounts for the thermodynamical cost. Details on their construction are reported in \sect{sec.fE}, \sect{sec.fQ} and \sect{sec:thermodynamicalcost} respectively.}
\end{table}

\subsection{Quantum Speed Limit  $g_\text{\tiny S}$}
\label{sec.fE}

We consider as the first quality the protocol time. Here, inspired from information processing, the shorter the better. Shorter timescales are also beneficial for having reduced interaction with the environment, and thus lower decoherence \cite{Goold_2016,Schlosshauer_2019}.
The time to implement the protocol has a fundamental lower bound determined by the quantum speed limit (QSL) \cite{Deffner_2017}, which imposes a minimum time $\tau_\text{\tiny QSL}$ to be able to distinguish two states during an evolution.
{For a closed dynamics with time-independent Hamiltonian, the QSL imposes the following bound on a quantum evolution \cite{Deffner_2017},
\begin{equation}\label{equ:quantumspeedlimit}
  \tau \geq \max\left\{ \frac{\hbar}{\Delta E_\tau} \mL(\rho_i,\rho_f), \frac{2\hbar}{\pi \braket{E}_\tau} \mL(\rho_i,\rho_f)^2 \right\},
\end{equation}
where $\braket{E}_\tau$ is the time-averaged mean energy , and $\Delta E_\tau$ is the time-averaged energy variance. $\mathcal L(\rho_i, \rho_f)$ is the Bures angle between the initial state $\rho_i$ and the final state $\rho_f$, and it is defined as
\begin{equation}
  \mL(\rho_i,\rho_f) = \arccos{(\sqrt{F(\rho_i,\rho_f)})},
\end{equation}
with $F(\rho_i,\rho_f)$ is the quantum fidelity, which reads
\begin{equation}
  F(\rho_i,\rho_f) = \tr{\sqrt{\sqrt{\rho_i}\rho_f\sqrt{\rho_i}}}^2.
\end{equation}
The Bures angle serves to quantify the distance between two states.
The two expressions compared in \eq{equ:quantumspeedlimit} are respectively the Mandelstam-Tamm limit \cite{Mandelstam_1991} and the Margolus-Levitin limit \cite{Margolus_1998}. It can be shown that the maximum between these limits provides the proper upper bound to the QSL \cite{Levitin_2009}. A generalization of the QSL in \eq{equ:quantumspeedlimit} has been introduced to generic open positive dynamics with time-dependent Hamiltonian \cite{Deffner_2013_2}, such that
\begin{equation}\label{equ:geometricquantumspeedlimit}
  \tau \geq \max\left\{ \frac{1}{\Lambda_\tau^\text{\tiny op}},\frac{1}{\Lambda_\tau^\text{\tiny hs}},\frac{1}{\Lambda_\tau^\text{\tiny tr}}\right\}\hbar\sin^2\left[\mathcal{L}(\rho_i,\rho_f)\right],
\end{equation}
where $\{\Lambda_\tau^\text{\tiny op},\Lambda_\tau^\text{\tiny hs},\Lambda_\tau^\text{\tiny tr}\}$ are the time-averaged operator, Hilbert-Schmidt and trace norms, respectively. Here, we decide to use the bound tightened by the Hilbert-Schmidt norm $\norm{A}_\text{\tiny hs} = \sqrt{\tr{A^\dag A}}$, which is linked to the Mandelstam-Tamm limit in closed dynamics,  due to its connection to the energetic cost associated with implementing the shorcut in Hamiltonian $H_1(t)$ \cite{Campbell_2017,Zheng_2016}, namely we take
\begin{equation}\label{equ:genericmtqsl}
    \tau_\text{\tiny QSL} = \frac{\hbar}{\Lambda_\tau^\text{\tiny hs}} \sin^2\left[\mathcal{L}(\rho_i,\rho_f)\right],
\end{equation}
where $\Lambda_\tau^\text{\tiny hs} = \frac{1}{\tau}\int_0^\tau\D t\norm{L[\rho_t]}_\text{\tiny hs}$ and $L[\rho_t]$ is the generator of the dynamics (this will be eventually the right-hand-side of \eq{equ:systemmasterequation}).}

Given that $\tau\geq\tau_\text{\tiny QSL}$, we quantify the speed $g_\text{\tiny S}$ with the following coarse-grained function,
\begin{equation}\label{eq.def.gs}
  g_\text{\tiny S} = \max \left\{ 0,\, \left(1-0.1\times\log_{10}\frac{\tau}{\tau_\text{\tiny QSL}}\right) \right\},
\end{equation}
which measures the magnitude of the ratio between $\tau$ and $\tau_\text{\tiny QSL}$. In particular, we make the speed quality decreases by $0.1$ when the ratio increases one order of magnitude. To make an explicit example, if $\tau/\tau_\text{\tiny QSL}\sim 1$ then $g_\text{\tiny S}\approx 1$; if $\tau/\tau_\text{\tiny QSL}\sim 10$ then $g_\text{\tiny S}\approx 0.9$, etc. The definition of $g_\text{\tiny S}$ we consider automatically sets $g_\text{\tiny S}$ to zero for any protocol requiring a time $\tau$ being equal or larger than $10^{10}\tau_\text{\tiny QSL}$. This can be considered a flaw of the definition. However, alternative definitions of $g_\text{\tiny S}$, such as $\tau_\text{\tiny QSL}/\tau$, would not give proper weight to really good protocols (e.g.~$\tau=10^2\tau_\text{\tiny QSL}$).
Notice that $\tau_\text{\tiny QSL}$ is determined by the generalised time-averaged energy variance [\ie~$\Lambda_\tau^\text{\tiny hs}$] and the distance between the initial and final states [\ie~$\sin^2\left[\mathcal{L}(\rho_i,\rho_f)\right]$]. This means that for an efficient protocol, one needs to have a small value of $\Lambda_\tau^\text{\tiny hs}\tau / \sin^2\left[\mathcal{L}(\rho_i,\rho_f)\right]$.

\subsection{Experimental quality $g_\text{\tiny Q}$}
\label{sec.fQ}

The goal of the protocol is to transfer a state from one to the other well of a double-well potential. The quality of the protocol is reflected by how much the final state $\rho_f$ after having run the protocol is near to the target state $\rho_\text{\tiny TG}$ of the protocol task. Thus, we choose to quantify the experimental quality $g_\text{\tiny Q}$ with the fidelity $F$ between these two states:
\begin{equation}
g_\text{\tiny Q} = F_\text{exp}(\rho_f,\rho_\text{\tiny TG}).  
\end{equation}
We add the subscript \texttt{exp} to stress that such a fidelity is also subject to experimental imperfections (such those imposed by the statistical and systematic errors), and performing these protocols in real experiments would usually result in states with lower fidelity compared to the theoretical one.  Here, we focus only on the later fidelity, which is given by the action of the protocol.

The protocol task and its target state $\rho_\text{\tiny TG}$ are defined as follows. We consider the double-well potential as divided in two local potentials, and we denote respectively with $\{\ket{j(t)}_\text{\tiny L}\}$ and $\{\ket{j(t)}_\text{\tiny R}\}$ the corresponding instantaneous eigenstates at time $t$ localised in the left and right well. We assume that the initial state of the problem is localised in the right well, and thus can be described as a linear superposition of the right instantaneous eigenstates, i.e.
\begin{equation}\label{equ:initialstate}
  \ket{\psi(0)}=\sum_j \alpha_j\ket{j(0)}_\text{\tiny R}.
\end{equation}
In the case where one does not apply the transfer protocol, such a state evolves as
\begin{equation}\label{equ:targetstateright}
  \ket{\psi(t)}=\sum_j \alpha_j e^{-i E^\text{\tiny R}_j t / \hbar} \ket{j(0)}_\text{\tiny R},
\end{equation}
where $E^\text{\tiny R}_j$ is the eigenvalues of $H_0(0)$ corresponding to the eigenstate $\ket{j(0)}_\text{\tiny R}$ in the right well. Now, this is the state one would like to have but transferred in the left well. Then, the target state is given by 
$$\rho_\text{\tiny TG} =\ket{\phi(\tau)}\bra{\phi(\tau)},
$$
where 
\begin{equation}\label{equ:targetstate}
  \ket{\phi(\tau)}=\sum_j \alpha_j e^{-i E^\text{\tiny R}_j \tau/\hbar} \ket{j(0)}_\text{\tiny L},
\end{equation}
where the relative weights $\alpha_j$ and the energies $E^\text{\tiny R}_j$ are the same as in Eq.~\eqref{equ:targetstateright}, but its decomposition is done on the left instantaneous eigenstates $\ket{j(0)}_\text{\tiny L}$. In such a way, the state $\ket{\phi(\tau)}$ in Eq.~\eqref{equ:targetstate} has the same information content of the state $\ket{\psi(t=\tau)}$ in Eq.~\eqref{equ:targetstateright}, and one can say that the state has been perfectly transferred.

\subsection{Thermodynamic cost $g_\text{\tiny T}$}\label{sec:thermodynamicalcost}
In experiments, the system is always under the influence of environment, and unavoidably undergoes to non-equilibrium processes if the protocol is performed in a finite time.
To account for such effects, we consider the following master equation
\begin{equation}\label{equ:systemmasterequation}
\dot{\rho} = -\frac{i}{\hbar} [H_\text{sys}, \rho] + D_\text{lc}[\rho] + D_\text{th}[\rho],
\end{equation}
where $D_\text{lc}[\rho]$ and $D_\text{th}[\rho]$ are respectively the localisation and thermal dissipators.
The first one describes the decoherence due to photon recoil \cite{Gonzalez-Ballestero_2019}, which has the form \cite{Joos_1985} 
\begin{equation}\label{equ:lcdissipator}
  D_\text{lc}[\rho] = - \Lambda[x,[x,\rho]],
\end{equation}
where $\Lambda$ is the corresponding diffusion constant. The thermal effects of the interactions with an environment with a temperature $T$ are instead accounted by $D_\text{th}[\rho]$, which is a complete positive version of Caldeira-Leggett dissipator reading \cite{Caldeira_1983,Gao_1997,Lampo_2016,Homa_2019}
\begin{equation}\label{equ:cldissipator}
\begin{aligned}
  D_\text{th}[\rho] =& - \frac{i \gamma}{2\hbar}[x,\{p,\rho\}] -\sum_{q=x,p}\gamma_q[q,[q,\rho]] 
\end{aligned}
\end{equation}
with $\gamma_x={\gamma m \kB T}/{\hbar^2}$ and $\gamma_p={\gamma}/(16 m \kB T)$. Here $\gamma$ is the coupling strength with the environment, $m$ is the mass of the particle, and $\kB$ is the Boltzmann constant.

The interaction with the environment  results in an irreversible entropy production $\Sigma_\text{ir}\geq 0$ \cite{Landi_2021}, which would be redeemed with exchanging heat between the system and the environment during the thermalisation. Usually this indicates that the process is no longer time-reversible and that the system information is lost.
To quantify such a thermodynamic cost, we use
\begin{equation}\label{eq.def.gt}
  g_\text{\tiny T} = e^{-\Sigma_\text{ir}},
\end{equation} 
which resembles the expression from fluctuation theorems \cite{Crooks_1999,Wang_2002,Jarzynski_2004,Funo_2018}, and it favors the process with small irreversible entropy production. 

By building on the results of our previous work \cite{Qiongyuan_2022}, 
we use the Wehrl entropy to measure the above quantity, which reads \cite{Santos_2017}
\begin{equation}\label{equ:wehrlentropy}
  S_\mQ = - \int \D \alpha\int \D \alpha^\ast\,\mQ_{\rho}(\alpha,\alpha^\ast) \ln \mQ_{\rho}(\alpha,\alpha^\ast),
\end{equation}
where $\mQ_{\rho}(\alpha,\alpha^\ast)$ is the Husimi Q-function being defined in terms of the coherent state $\ket{\alpha}$:
\begin{equation}\label{husimi.def}
  \mQ_{\rho}(\alpha,\alpha^\ast) = \frac{1}{\pi}\bra{\alpha}\rho\ket{\alpha}.
\end{equation}
In the case under study, one does not have a harmonic potential and thus the coherent states $\ket \alpha$ are more difficult to define. Nevertheless, for low energies, one can approximate the single well as an harmonic oscillator with frequency $\omega=(E_1-E_0)/\hbar$, where $E_0$ and $E_1$ are respectively the ground and first-excited state energies of one of the two wells. In such a way, one can define the creation and annihilation operators $a^\dag$ and $a$ through $x=\sqrt{\frac{\hbar}{2 m \omega}}(a+a^\dagger)$, $p=-i\sqrt{\frac{\hbar m \omega}{2}}(a-a^\dagger)$; and the corresponding coherent states $\ket \alpha$ follow straightforwardly. We note that
 the final result for the irreversible entropy production does not dependent on the choice of $\omega$, underlying the strength of this reasoning.
 
Conversely to the von Neumann entropy, the Wehrl entropy has well-defined decomposed rates at the zero temperature limit, and is the upper-bound to the von Neumann entropy \cite{Abe_2003,Audenaert_2014,Santos_2017,Santos_2018}. To compute the irreversible entropy production $\Sigma_\text{ir}$, we first take the time derivative of the Wehrl entropy. Given the master equation in \eq{equ:systemmasterequation}, we can decompose the latter as
\begin{equation}
  \frac{\D S_\mQ}{\D t} = \frac{\D S_\mU}{\D t} + \frac{\D S_{\mD_\text{lc}}}{\D t} + \frac{\D S_{\mD_\text{th}}}{\D t},
\end{equation}
where the first term is the Wehrl entropy rate for the unitary process, and the two other terms are the rates for the dissipative processes. In general, one can decompose the rate of a dissipative process as 
\begin{equation}
  \frac{\D S_\mD}{\D t} = \Pi - \Phi,
\end{equation}
where the first term $\Pi$ is the irreversible entropy production rate, and the second term $\Phi$ is the entropy flux rate. It follows that the total irreversible entropy production during the process is 
\begin{equation}\label{equ:ratetoproduction}
  \Sigma_\text{ir} = \int_0^\tau \D t\,\Pi(t).
\end{equation}
Here, the irreversible entropy production rate $\Pi$ comes from the component that is even to the time reversal of the entropy rate, and the entropy flux rate $\Phi$ comes from the component that is odd \cite{Brunelli_2016}. It has been shown \cite{Santos_2018} that for a dissipator of the form
\begin{equation}\label{eq.OO}
  D[\rho] = -[O,[O,\rho]],
\end{equation}
with $O=x$ or $p$, one has solely the irreversible entropy production rate, which is given by
\begin{equation}
\Pi = \int \D \alpha\int \D \alpha^\ast\, \frac{\abs{\mC_O(\mQ)}^2}{\mQ},
\end{equation}
and depends on the even power of the current $\mC_O(\mQ)=\braket{\alpha|[O,\rho]|\alpha}$.
In this way, one can account for the dissipator $D_\text{lc}[\rho]$ and the last two terms of $D_\text{th}[\rho]$ as defined in \eq{equ:cldissipator}. On the other hand, the first term of $D_\text{th}[\rho]$ needs to be tackled in a different way. 
We show in App.~\ref{apd:decomposerates} that it can be decomposed in two parts as
\begin{equation}\label{equ:decomposerates}
 - \frac{i \gamma}{2\hbar} [x,\{p,\rho\}] 
    =-\frac{\gamma}{2\hbar}[x,[x,\rho]] -\frac{\gamma}{\sqrt{2}\hbar}[x,\rho a - a^\dag\rho].
\end{equation}
In particular, the first part has the form appearing in Eq.~\eqref{eq.OO} and thus it contributes to the irreversible entropy production rate $\Pi$ only. The second part contributes instead to the entropy flux rate, since it corresponds a contribution to the entropy rate $\Phi$ that is odd in the current [cf.~App.~\ref{apd:decomposerates}].
Therefore, the explicit expressions of the total irreversible entropy production and the flux rates of the model in \eq{equ:systemmasterequation} are
\begin{equation}
  \begin{aligned}
    \Pi &= \left( \frac{\gamma}{2\hbar} +\gamma_x  
    + \Lambda \right) \int \D\alpha\int \D\alpha^\ast\, \frac{\abs{\mC_x(\mQ)}^2}{\mQ} \\
    &\quad+ \gamma_p 
    \int \D\alpha\int \D\alpha^\ast\, \frac{\abs{\mC_p(\mQ)}^2}{\mQ},\label{equ:irreversibleentropyproduction} \\
    \Phi &=\frac{\gamma}{\sqrt{2}\hbar} \int \D\alpha \int \D\alpha^\ast\, |\alpha|^2 \, \mC_x(Q),
  \end{aligned}
\end{equation}
from which one can calculate the irreversible entropy production through \eq{equ:ratetoproduction}.

\section{Protocols for the state transfer}\label{sec:protocols}

\begin{figure}[t]
  \centering
  \includegraphics[width=0.8\linewidth]{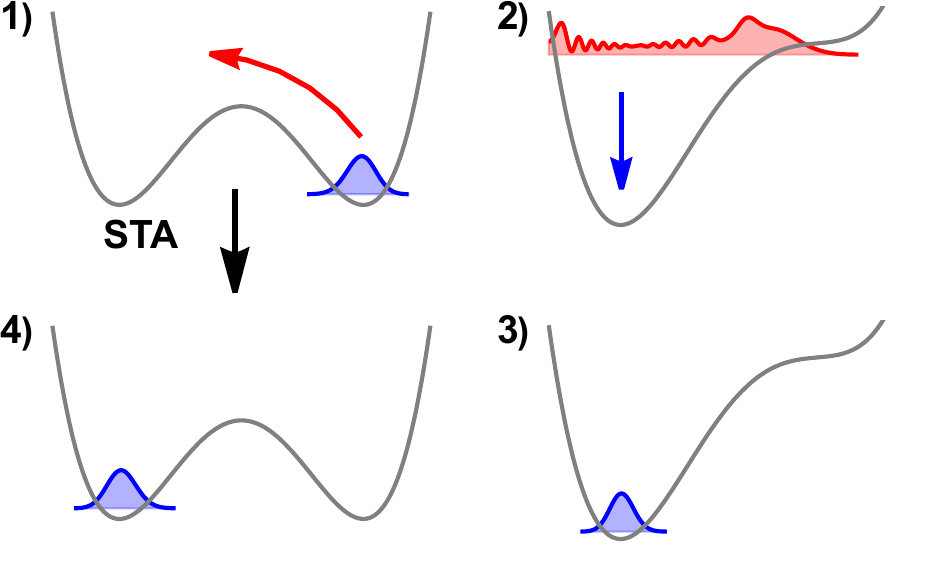}
  \caption{Schematic illustration of the classical and the quantum state transfer protocols. The classical protocol follows the frame sequence $\mathbf{ 1)\to2)\to3)\to4)}$, while the quantum protocol goes directly from $\mathbf{ 1)\to4)}$ with the help of STA. 
  }
  \label{fig:protocolscheme}
\end{figure}

\begin{figure*}[t]
\stackunder[5pt]{\textbf{(a)}}{\includegraphics[width=0.32\textwidth]{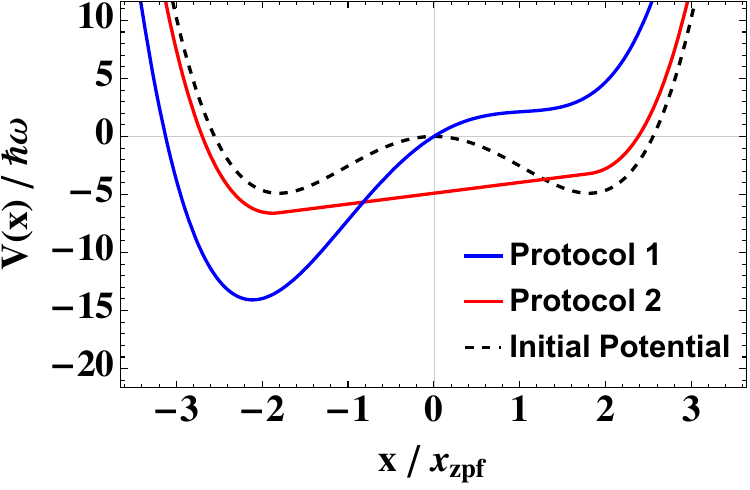}}\hfill
\stackunder[5pt]{\textbf{(b)}}{\includegraphics[width=0.32\textwidth]{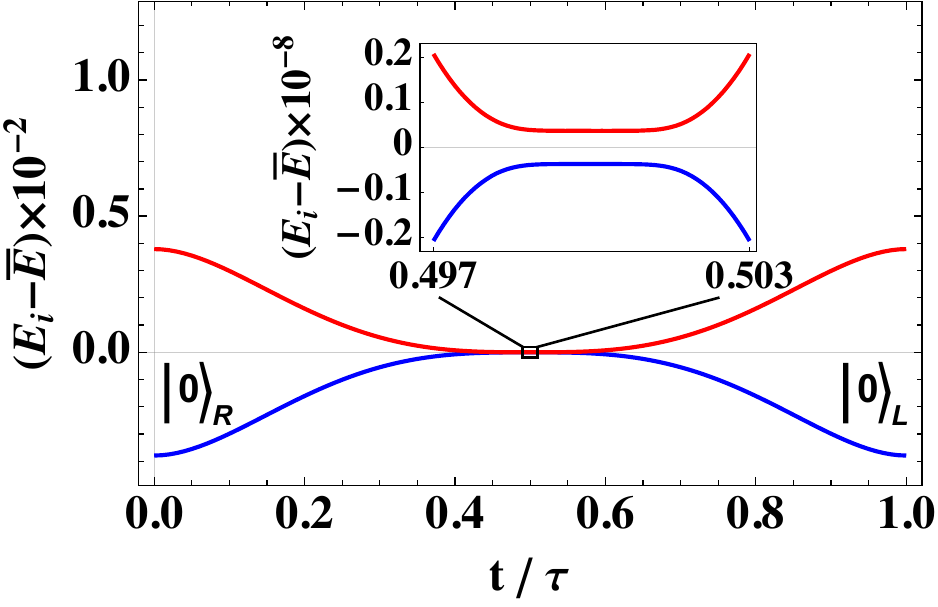}}\hfill
\stackunder[5pt]{\textbf{(c)}}{\includegraphics[width=0.32\textwidth]{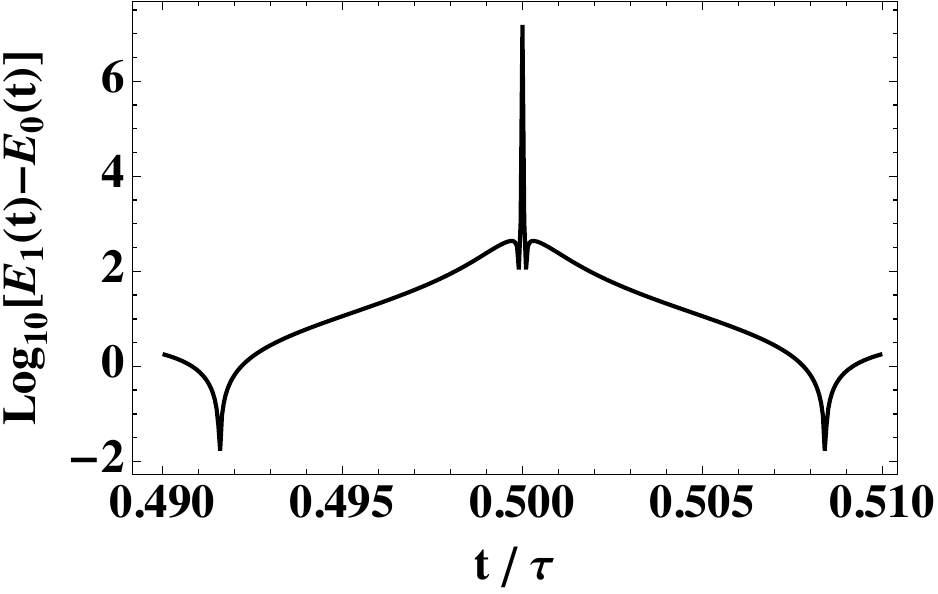}}
\caption{Panel \textbf{(a)} shows the shapes of the trapping potentials for the two classical protocols implemented with $f_\text{1}^\text{\tiny C} (x,t)$ and $f_\text{2}^\text{\tiny C} (x,t)$, respectively shown as blue and red lines, compared to the initial potential at $t=0$ represented with a black dashed line. Here, we set $x_\text{\tiny zpf} = \sqrt{\hbar/2m\omega}$, with $\omega \approx 2.3$, $c_1=-1.5$, $c_2=0.05$, $\delta=0.001$, and $a_\text{1}^\text{\tiny C}=5$, $a_\text{2}^\text{\tiny C}=b_\text{2}^\text{\tiny C}=1$; the time $t$ is set to that maximising the change in the potential. 
Panel \textbf{(b)} shows the dynamics of the ground (blue line) and first-excited (red line) energies of $H_0(t)$ with the control parameter $f_\text{2}^\text{\tiny Q}(x,t)$, whose value changes from $-\delta x$ to $\delta x$ over time window $t\in[0,\tau]$. As one can see their distance is of the order of $10^{-9}$ in the central part of the protocol (see inset). Panel \textbf{(c)} shows the difference between the first-excited and ground state when applying the STA protocol (in particular that with the control parameter described in \eq{eq.linearf} below). The energy difference in the center of the protocol  ($t=\tau/2$) is now big $\sim 10^7$. Two minima of such a difference appear on the sides of the center with a value of the order of $10^{-2}$, which is still seven orders of magnitude larger than the absolute minimum of the energy difference in the classical protocol.}
	\label{fig:sketchesofprotocols}
\end{figure*}

In order to showcase the use of the protocol grading that we defined in \sect{sec:quantifier}, here we consider the specific problem of transferring a state from one side to the other of a double-well potential. We consider an one-dimensional system in a double-well potential, whose corresponding Hamiltonian reads
\begin{equation}\label{equ:doublewellsystemfree}
  H_\text{free} = \frac{p^2}{2m} + c_1 x^2 + c_2 x^4 ,
\end{equation}
where the coefficients $c_1<0$ and $ c_2>0$ determine the shape of the double-well potential.
Next, we will introduce two particular classical protocols and two quantum ones for the state transfer. The classical protocols only deform the potential, and the state must go over the barrier to reach the other side. Conversely, the quantum protocols harness genuine quantum effects, namely the quantum tunnelling, such that the state transfers through the barrier between the two wells with the help of the counterdiabatic term. A schematic illustration of the protocols is shown in \fig{fig:protocolscheme}.  

\subsection{Classical state transfer}

In order to switch the state of a trapped particle form the left to the right well, we allow an external agent to modify the potential through an external control term that is added up to $H_\text{free}$ as
\begin{equation}\label{equ:doublewellsystem}
  H_0(t) = H_\text{free} + f(x,t).
\end{equation}
The role of function of $f(x,t)$ is to deform the potential. We consider two classical transfer protocols, which correspond to the following  control functions
\begin{align}
  f_1^\text{\tiny C} (x,t) &= \alpha^\text{\tiny C}_1(t)x,\label{equ:classicalcontrol1}\\[2ex]
  f_2^\text{\tiny C} (x,t) &= \alpha^\text{\tiny C}_2(t)x-\beta^\text{\tiny C}_2(t)
    \left(\frac{c_1^2}{4c_2}+c_1 x^2 + c_2 x^4\right)\notag\\
    &\times\theta\left(-\frac{c_1}{2c_2}-x^2\right) . \label{equ:classicalcontrol2}
\end{align}
The first control function simply tilts the potential by adding a linear term $\alpha^\text{\tiny C}_1(t)x$ to the potential. We choose the parameter $\alpha^\text{\tiny C}_1(t)$ to change linearly in time from $-\delta<0$ to $A_1>0$. Here, the positive value of $\delta$ is small but non-zero which is needed to make the initial potential asymmetric (slightly tilted towards the right side), so that the ground state of the system is initially localised in the right well. The value of $A_1$ determines the degree of the maximal tilting.
The second considered control function, in addition to the linear tilting process controlled by $\alpha^\text{\tiny C}_2(t)\in[-\delta,A_2]$,
adds another term that linearly flattens the central part of the double-well potential (namely the wall between the two wells), and it is controlled by $\beta^\text{\tiny C}_2(t)\in[0,1]$. Such a choice has already been experimentally realised with trapped underdamped nano- and microparticle in the classical, stochastic regime  \cite{Berut_2012,Hong_2016,Ciampini_2021}.
These two protocols implement the transfer in two structurally different ways. In short, the first classical protocol just tilts the potential, while the second one tilts it and flattens its central part. In \subf{fig:sketchesofprotocols}{a}, we
compare the two potentials at the time of the corresponding strongest imposed tilting.

The first classical protocol (dubbed {\it classical protocol 1}) is illustrated in  \fig{fig:protocolscheme} with the frame sequence $\mathbf{ 1)\to2)\to3)\to4)}$ and it is implemented as follows. We assume that the system is initially localised in the ground state in the right well [cf.~frame {\bf 1)}] and we want to transfer it to the ground state of the left well [cf. frame {\bf 4)}]. 
When acting on the system with the control function $f_1(x,t)$, the energy of the system unavoidably grows due to the non-adiabatic deformation of the potential, and thus high-energy states of system get populated [cf.~frame {\bf 2)}]. Since the high-energy wavefunctions extend over the entire potential, the position of the system is no longer localised, as it is shown by the red wavepacket in frame {\bf 2)}. To drive the system back to the ground state (which is now in the left well) [cf.~frame {\bf 3)}], we need to cool down the system. This can be done by attaching it to an environment at low temperatures, so that after some time the system will dissipate heat to the environment. In particular, we attach the latter to the system already 
at the beginning of the protocol at time $t=0$.
In general, the energy increase is large when a fast tilting is performed, while in an adiabatic (infinite time) process it is smaller although lower-bounded by a generalized Landauer principle \cite{Landauer_1961,Berut_2012,Konopik_2020}. A fast tilting process is usually not desired due to the large energy increase and the consequent long-time cooling needed. Finally, one can reverse the protocol, namely run the control functions backwards, and restore the initial potential [cf.~frame {\bf 4)}], which completes the classical transfer protocol. 

The second classical protocol (which we refer to as {\it classical protocol 2}) works in a similar manner as the first one. Again, the frame sequence the system will follow is given by $\mathbf{ 1)\to2)\to3)\to4)}$, with the only difference being that the potential is not only tilted but also flattened in its central part (see for instance the red line in \subf{fig:sketchesofprotocols}{a}). Again, there will be a heating of the system due to the tilting and flattening, which will lead to a spread of the wavepacket and it will require a cooling process.

\subsection{Quantum state transfer}

To construct the quantum state transfer protocol, we start from the Hamiltonian $H_0(t)$ considered for the classical protocol in \eq{equ:doublewellsystem}.
For the sake of simplicity, here we consider the application of STA only to the first classical protocol. Indeed, STA does not apply well to the second protocol
due to the degeneracy between the energies of the left and right well appearing when the potential is flatten \cite{Campbell_2015,Guery-Odelin_2019}.

The quantum part of classical protocol 1 is implemented through the assistance of STA. In particular, a counter-diabatic (CD) driving such the one defined in Refs.~\cite{Berry_2009,Guery-Odelin_2019} is introduced through the following:
\begin{equation}\label{equ:STAfullH}
  H_1(t) = H_0(t) + H_\text{\tiny STA}(t),
\end{equation}
where we included the counter-diabatic term, which reads $H_\text{\tiny STA}(t) = i\hbar \sum_{i}\,\bigl[\ket{\partial_ti}\bra{i},\ket{i}\bra{i}\bigr]$, where $\ket{i}=\ket{i(t)}$ are the instantaneous eigenstates of the Hamiltonian $H_0$ with corresponding energies $E_i=E_i(t)$, such that $H_0(t)\ket{i(t)} = E_i(t) \ket{i(t)}$. The explicit form of $\ket{\partial_t i}$ can be computed with the Schr\"odinger equation \cite{Berry_2009} and thus $H_\text{\tiny STA}$ can be recast as 
\begin{equation}\label{equ:cdterm} 
  H_\text{\tiny STA}(t) =i\hbar\sum_{i\neq j}\frac{\bra{i}\dot{H_0}\ket{j}}{E_j-E_i}\ket{i}\bra{j}.
\end{equation}
In order to ensure that the initial and final potentials are those of the original Hamiltonian $H_0$, we impose the STA conditions $H_\text{\tiny STA}(0)=H_\text{\tiny STA}(\tau)=0$ on the CD term. 

The quantum protocol we consider here is similar to that in Ref.~\cite{Gaudenzi_2018}, where the protocol exploits the tunnelling effect in a double-well potential. Here, we adopt the STA to achieve and accelerate the state transfer. In particular, the protocol works due to the following reasons: \textit{i)} The sign change of the control parameter $f(x,t)\in[-\delta x,\delta x]$ flips the asymmetry of the potential in \eq{equ:doublewellsystem}, switching the ground state from being in right well to the left well (see a simple application to the Landau-Zener model in Appendix \ref{apd:landauzenermodel}); and \textit{ii)} the solutions of the new Hamiltonian $H_1(t)$ follow, in finite time $\tau$, the adiabatic trajectory of the original Hamiltonian $H_0$. In \subf{fig:sketchesofprotocols}{b}, we plot the dynamics of the eigenvalues for the ground state (in blue) and first-excited state (in red) of $H_0$. In the classical protocol, when starting from the ground state in the right well $\ket{0}_\text{\tiny R}$ and move towards that in the left $\ket{0}_\text{\tiny L}$, due to the energy increase, the system will jump from the ground state to the first-excited one (i.e., from the blue to the red line) when the energy gap is small enough (this happens at $t/\tau=0.5$). To prevent this, the quantum protocol increases the energy gap between the ground and the first-excited state, as depicted in \subf{fig:sketchesofprotocols}{c}. In such a way, 
it is more difficult to excite the system \cite{Campbell_2017}, at the cost of extra energy input given by the CD term. Therefore, the quantum protocol works as follows: we prepare the system in ground state in the right well $\ket{0}_\text{\tiny R}$, then the system evolves with the new Hamiltonian $H_1(t)$ and will follow the blue line, resulting in the ground state in the left well $\ket{0}_\text{\tiny L}$ (as depicted by the
frame sequence ${\bf1)}\to{\bf4)}$ in  \fig{fig:protocolscheme}).

We start with the system prepared in the ground state in the right well with the control parameter set to $f_i^\text{\tiny Q}(x,0)=-\delta x$.
{The state transfer is performed with the new Hamiltonian $H_1(t)$ and modifying the control parameter to a positive value $f_i^\text{\tiny Q}(x,\tau)=\delta x$. The way how $f_i^\text{\tiny Q}(x,t)$ is changed is determined by the STA conditions. Here we consider the control parameter to be proportional to the position operator as in \eq{equ:classicalcontrol1}:
\begin{equation}
  f_i^\text{\tiny Q}(x,t) = \alpha^\text{\tiny Q}_i(t)x.
\end{equation}
The substitution of the latter expression  in \eq{equ:doublewellsystem} gives $\dot{H}_0=\dot \alpha_i(t) x$ which is then merged with \eq{equ:cdterm}. Eventually, $H_\text{\tiny STA}(t)$ will be computed numerically [cf.~Sec.~\ref{sec:numericalsimulation}].
To the best of our knowledge, an analytical expression of such counterdiabatic term in a tilting double-well potential is still unknown, while preliminary studies to attain its analytical expression in a similar double-well potential can be found in Refs. \cite{Patra_2017, Patra_2021}. The STA conditions impose $\dot{\alpha}_i(0)=\dot{\alpha}_i(\tau)=0$. This considered, one can construct different interpolation of the control parameter connecting $\alpha^\text{\tiny Q}_i(0)=-\delta$ to $\alpha^\text{\tiny Q}_i(\tau)=\delta$}. Indeed, it has been shown that the energy cost of the counter-diabatic term $H_\text{\tiny STA}$ can be reduced by optimising the control parameter \cite{Abah_2019}. The first option we consider is a control parameter that is cubic in time and it is described by
\begin{equation}\label{eq.linearf}
    \alpha_1^\text{\tiny Q}(t)= -\delta + \frac{6\delta t^2}{\tau^2} 
  - \frac{4 \delta t^3}{\tau^3}.
\end{equation}
We indicate this control scheme to be linear as it grows linearly at $t=\tau/2$, this is when the energy gaps of $H_0(t)$ are the smallest (\eg~when the potential of $H_0$ is symmetric, and the denominator in \eq{equ:cdterm} is smallest, making at that time the contribution of $H_{\text{\tiny STA}}$ being the largest). 
In the second quantum protocol, we further require that $\dot{\alpha}_2^\text{\tiny Q}(\tau/2)=0$. In contrast with the linear interpolation in Eq.~\eqref{eq.linearf}, this gives the following non-linear behavior of the control parameter
\begin{equation}\label{eq.nonlinearf}
    \alpha_2^\text{\tiny Q}(t)= -\delta +\frac{ 30\delta t^2}{\tau^2}-\frac{100\delta t^3}{\tau^3}+\frac{120\delta t^4}{\tau^4}-\frac{48\delta t^5}{\tau^5} .
\end{equation}

\begin{figure}
  \centering
  \includegraphics[width=0.8\linewidth]{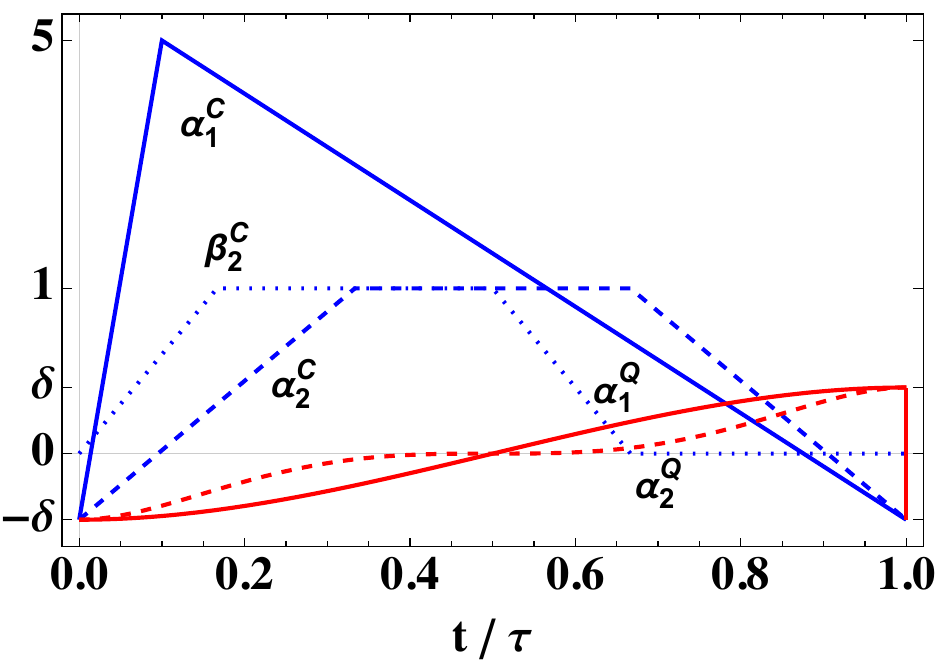}
  \caption{Schematic illustration of changes of all the functions in the proposed control parameters $f_i^k(x,t)$ with $i=1,2$, for the classical $k=\,$C and quantum $k=\,$Q protocols. 
  }
  \label{fig:controlparameters}
\end{figure}

In summary, we have introduced two options for classical state transfer, namely $f_1^\text{\tiny C}(x,t)$ and $f_2^\text{\tiny C}(x,t)$, and two options for quantum state transfer $f_1^\text{\tiny Q}(x,t)$ and $f_2^\text{\tiny Q}(x,t)$. In particular, $f_1^\text{\tiny C}(x,t)$ and $f_1^\text{\tiny Q}(x,t)$ are the simple linear protocols, while $f_2^\text{\tiny C}(x,t)$ and $f_2^\text{\tiny Q}(x,t)$ are non-linear protocols. We show a sketch of the changes for all control parameters in \fig{fig:controlparameters}.

\section{Numerical simulations of two protocols}\label{sec:numericalsimulation}

In this section, we lay down the basis for corroborating the theoretical framework with numerical simulations, whose basis was described in Ref.~\cite{Qiongyuan_2022}. As a case study, we set $c_1=-1.5$ and $ c_2=0.05$ (see \subf{fig:sketchesofprotocols}{a}), where the energy difference between the ground and first excited states in the right well is $\hbar\omega\approx2.3$, and let $m=\kB=\hbar=1$. We choose such a potential so that the energetic barrier is high enough to prepare well-localised states in either of the two wells, while still being low enough to highlight the features of the quantum protocol. Tunable potential landscapes of comparable energy scales have been realised in semiconductor qubits \cite{Chatterjee21}, as well as proposed in nanomechanical resonators coupled with quantum dots \cite{Pistolesi21}. Recent experiments with optically levitated nanoparticles provided the necessary toolbox for implementing time-controlled protocols in the quantum regime, as well as quantum-limited position readout and quantum initial state preparation \cite{Magrini21,Tebbenjohanns21}.  We choose to introduce the following numerical values for the coupling parameters $\gamma/\omega=10^{-2}$ and $\Lambda/\omega=10^{-3}$ from the master equation in Eq.~\eqref{equ:systemmasterequation}, following typical values in such experiments \cite{Qiongyuan_2022}. With this choice of the parameters, the energy at the top of barrier corresponds to a 
thermal state with associated temperature $T\approx12.7$\,K. Thus we considered two temperatures, $T=1$\,K and 10\,K, where the corresponding equilibrium systems have energies below the well potential and are well or loosely localised respectively. Finally, we consider for all protocols the initial state to be
\begin{equation}\label{eq.initialstate}
  \ket{\psi(0)} = 0.6\ket{0}_\text{\tiny R}+0.8\ket{1}_\text{\tiny R},
\end{equation}
correspondingly the target state at $t=\tau$ given by \eq{equ:targetstate} reads
\begin{equation}\label{eq.targetstatesimulation}
  \ket{\phi(\tau)} = 0.6 e^{-iE_0^R \tau /\hbar}\ket{0}_\text{\tiny L}+0.8 e^{-iE_1^R \tau /\hbar} \ket{1}_\text{\tiny L}.
\end{equation}

For the simulations, we use the full toolbox developed in \cite{Qiongyuan_2022}, which is concisely summarised below. The continuous system described by the bosonic field operators $\{a, a^\dag\}$ can be approximated by a discrete spin-$j$ system with ladder operators $\{J_+,J_-\}$ following the Holstein-Primakoff (HP) transformation, which can be stated as:
\begin{equation}
  a\approx M_\kappa^{-1}J_+,\quad\text{and} \quad a^\dag\approx J_-M_\kappa^{-1},
\end{equation}
where $M_\kappa$ is the $\kappa$-th order Taylor expansion of the nonlinear term in HP transformation,
\begin{equation}
  \hbar\sqrt{2j-a^\dag a}=M_\kappa + \mathcal{O}((a^\dag a)^{\kappa+1}).
\end{equation}
One can define the discretized version of dimensionless quadrature field operators as \cite{Weedbrook_2012}
\begin{equation}
  \begin{aligned}
    x'&=\frac{1}{\sqrt{2}}(J_-M_\kappa^{-1} + M_\kappa^{-1}J_+), \\
    p'&=\frac{i}{\sqrt{2}}(J_-M_\kappa^{-1} - M_\kappa^{-1}J_+),
  \end{aligned}
\end{equation}
and we have the discretized Hamiltonian, 
\begin{equation}\label{equ:discretedoublewellsystem}
  H(t)= \frac{{p'}^2}{2m} + c_1 {x'}^2 + c_2 {x'}^4 + f_\text{\tiny i}^\text{\tiny k}(x',t).
\end{equation}
and the control term can be chosen from $\{f_\text{\tiny i}^\text{\tiny k}(x',t)\}_{i=1,2}^\text{\tiny {k=C,Q}}$ depending on the specific protocol.
Here, the discrete system reflects the low-energy sector of the original continuous system in \eq{equ:doublewellsystem}, which is the sector we focus on. We set the dimension of the system $j$, as well as the Taylor expansion size $\kappa$, to be $60$. Namely, we have spin-$\frac{59}{2}$.

\begin{figure*}[tb]
\stackunder[5pt]{\textbf{(a)}}{\includegraphics[width=0.32\textwidth]{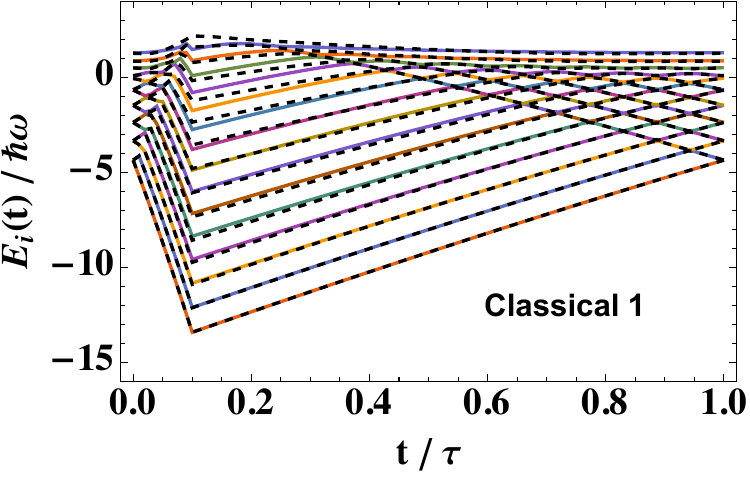}}\hfill
\stackunder[5pt]{\textbf{(b)}}{\includegraphics[width=0.32\textwidth]{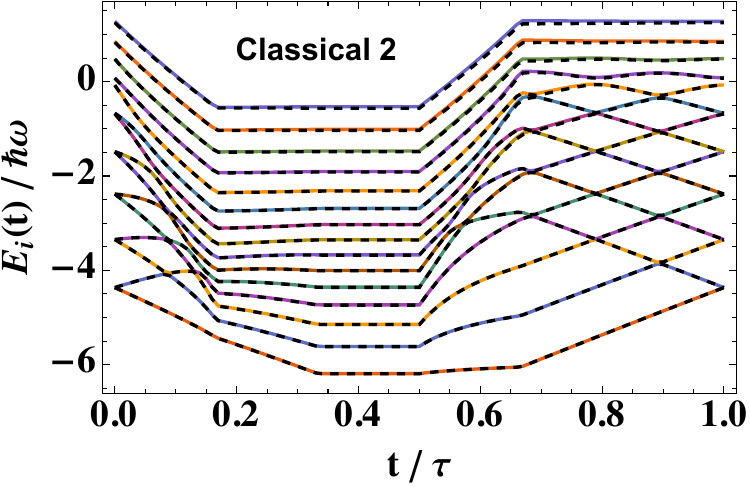}}\hfill
\stackunder[5pt]{\textbf{(c)}}{\includegraphics[width=0.32\textwidth]{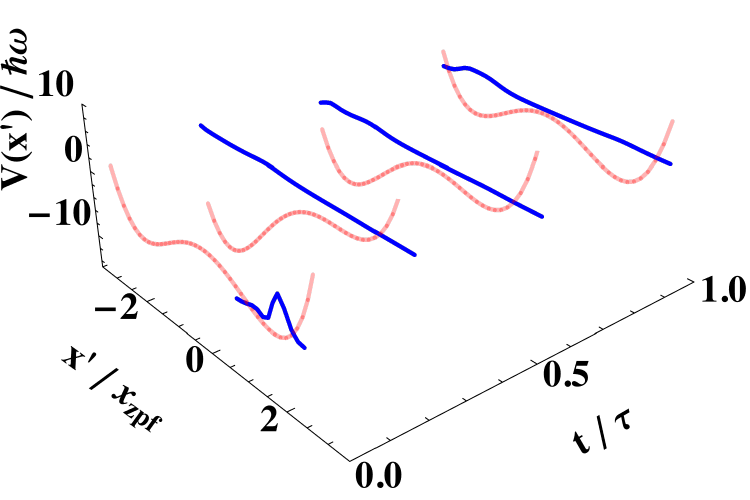}}\\\vspace{1em}
  \stackunder[5pt]{\textbf{(d)}}{\includegraphics[width=0.31\textwidth]{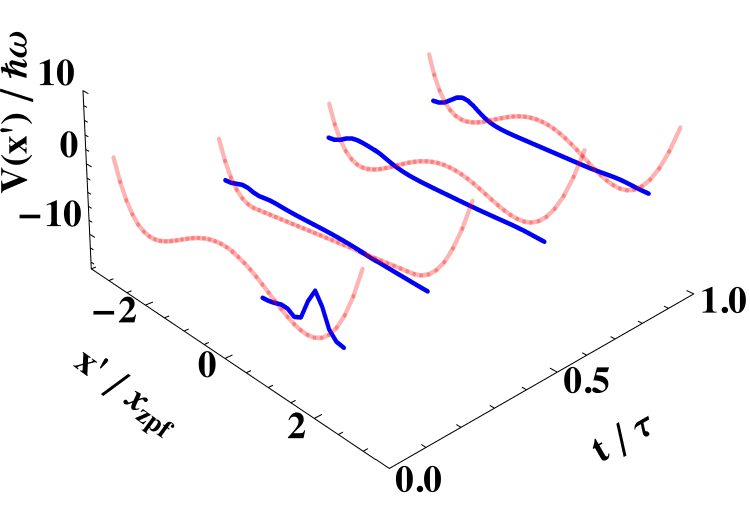}}\hfill
\stackunder[5pt]{\textbf{(e)}}{\includegraphics[width=0.32\textwidth]{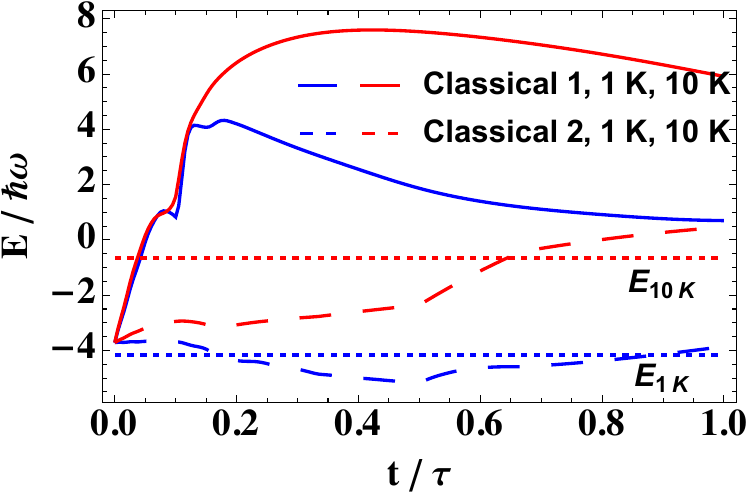}}\hfill
\stackunder[5pt]{\textbf{(f)}}{\includegraphics[width=0.34\textwidth]{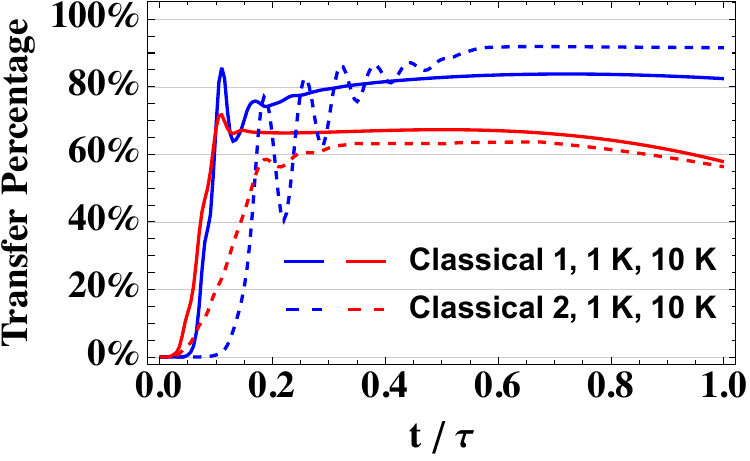}}
	\caption{Classical state transfer protocols. Panels \textbf{(a)} and \textbf{(b)} show the accuracy of the simulation, by comparing the evolution of the energy levels of the system for the two protocols computed with the continuous (colored lines) and with the discretised (dashed black lines) system. Panel \textbf{(c)} and panel \textbf{(d)} show the dynamics of the system's density in position (blue line) as the potential (red line) is changed in time at the temperature of 1\,K, respectively for the classical protocols 1 and 2. The main visible difference between the two panels is in the form of the potential for $t\sim \tau/3$.
	Panel \textbf{(e)} reports the system energy $\braket{H_0(t)}$ evolution when the system interacts with the environment of different temperatures (1\,K in blue and 10\,K in red). The continuous and dashed line correspond to protocol 1 and 2 respectively; the dotted lines correspond to the thermal energies and serve as a reference. Panel \textbf{(f)} shows the transfer percentage [cf.~\eq{eq.transferperc}] of the system state with respect to the corresponding thermal states. We employed the same color and dashing as in panel (e).}
	\label{fig:classicalunitary}
\end{figure*}

On the other hand, the coherent state $\ket{\alpha}$ in continuous system can be replaced with the spin-coherent state for discrete system \cite{Santos_2018},
\begin{equation}
  \ket{\Omega} = e^{- i \phi J_z} e^{-i \theta J_y} \ket{j,j},
\end{equation}
where $\ket{j,j}$ is the angular momentum state with largest quantum number of $J_z$, and $\Omega=\{\theta, \phi \}$ is the set of Euler angles identifying the direction along which the coherent state points. The corresponding Husimi-Q function is defined as
\begin{equation}
  \mQ(\Omega) = \bra{\Omega}\rho\ket{\Omega},
\end{equation}
the Wehrl entropy for a system $N=2j+1$ degrees of freedom reads
\begin{equation}
  S_\mQ = -\frac{N}{4\pi} \int \D\Omega\,\mQ(\Omega) \ln \mQ(\Omega),
\end{equation}
and the irreversible entropy production rate in \eq{equ:irreversibleentropyproduction} becomes
\begin{equation}\label{equ:irreversibleentropyproductionspin}
  \begin{aligned}
    \Pi =& \left( \frac{\gamma}{2\hbar} + \frac{\gamma m \kB T}{\hbar^2} + \Lambda \right) \frac{N}{4\pi} \int \D\Omega\, \frac{\abs{\mJ_x(\mQ)}^2}{\mQ} \\
     &+ \frac{\gamma}{16 m \kB T} \frac{N}{4\pi} \int \D\Omega\, \frac{\abs{\mJ_p(\mQ)}^2}{\mQ},
  \end{aligned}
\end{equation}
where $\mJ_O(\mQ)=\bra{\Omega}[O,\rho]\ket{\Omega}$ with $ O=x, p$.

\subsection{Classical protocol}
\label{sec.classicalprotocol}

The simulations for the classical protocols are performed by employing the Hamiltonian in Eq.~\eqref{equ:doublewellsystem} with two forms of the control parameter: for the protocol 1 we use $f_1^\text{\tiny C}$ defined in Eq.~\eqref{equ:classicalcontrol1}, while for protocol 2 we employ $f_2^\text{\tiny C}$ presented in Eq.~\eqref{equ:classicalcontrol2}. The control parameters take the values $\delta=0.001$, $ A_1=5$, $ A_2=1$, and we set $\tau\omega=300$. As $A_2 < A_1$, protocol 2 tilts the potential less than protocol 1. Indeed, in the  protocol 2, the central barrier is also lowered, thus the protocol excites less to the system and because of this is the optimised one.

\begin{figure*}[t]
    \stackunder[5pt]{\textbf{(a)}}{\includegraphics[width=0.32\textwidth]{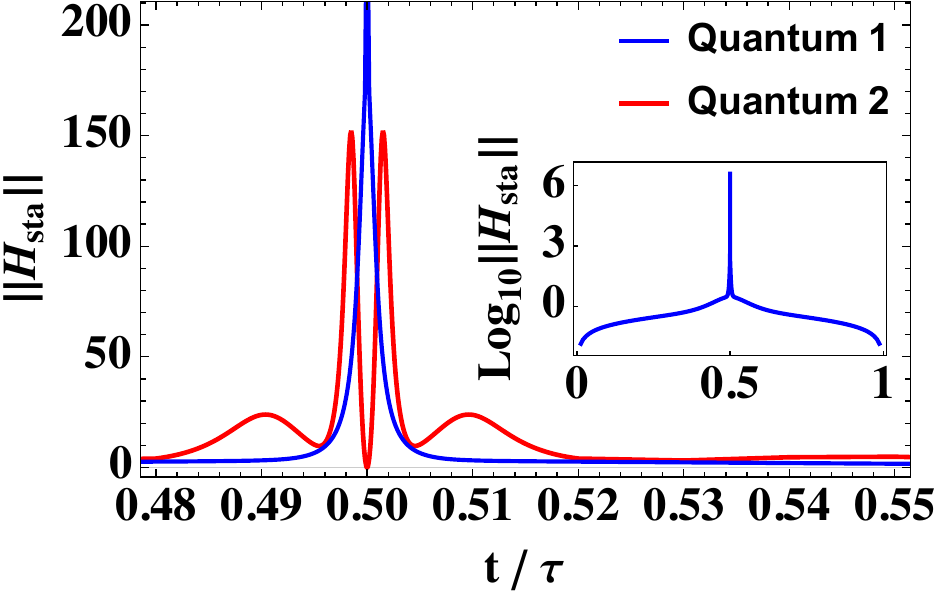}}\hfill
    \stackunder[5pt]{\textbf{(b)}}{\includegraphics[width=0.32\textwidth]{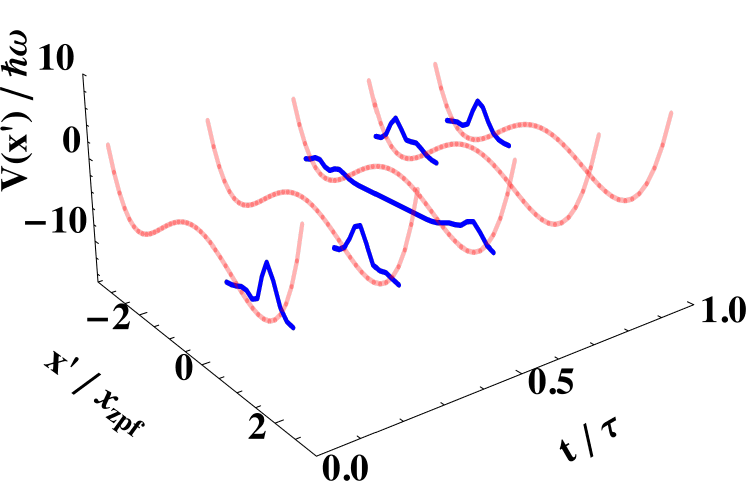}}\hfill
    \stackunder[5pt]{\textbf{(c)}}{\includegraphics[width=0.32\textwidth]{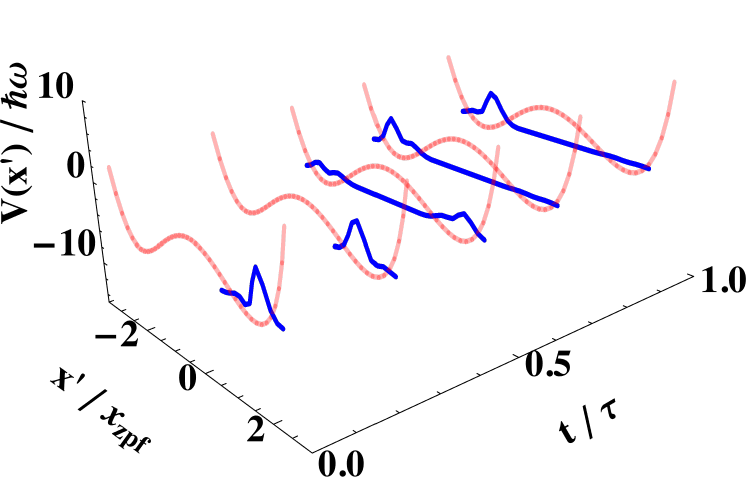}}\\\vspace{1em}
    \stackunder[5pt]{\textbf{(d)}}{\includegraphics[width=0.31\textwidth]{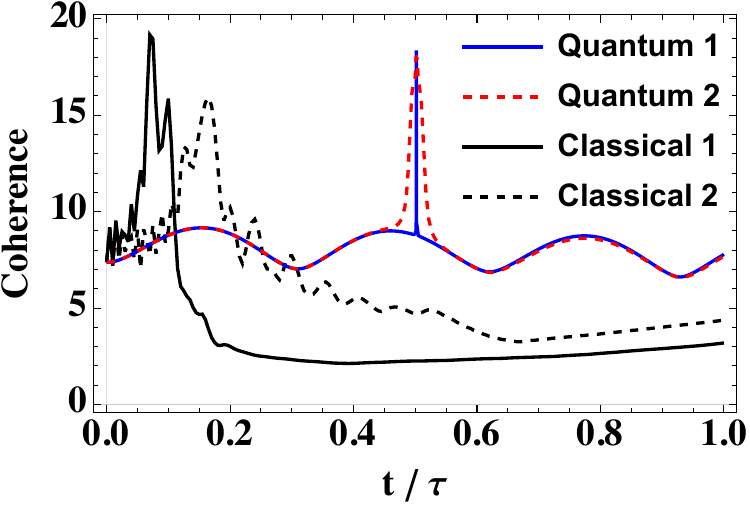}}\hfill
    \stackunder[5pt]{\textbf{(e)}}{\includegraphics[width=0.32\textwidth]{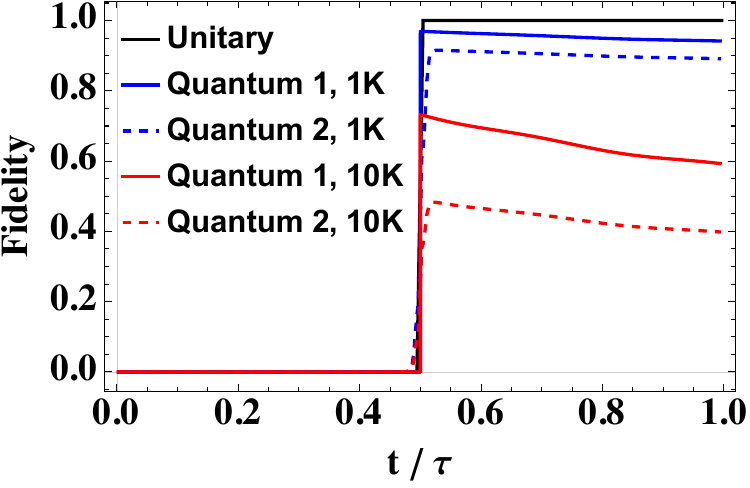}}\hfill
    \stackunder[5pt]{\textbf{(f)}}{\includegraphics[width=0.34\textwidth]{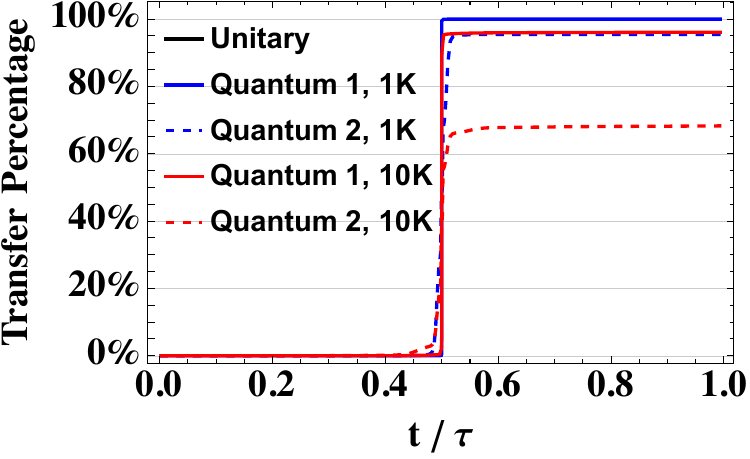}}
	\caption{The quantum state transfer protocol.  Panel \textbf{(a)} shows the trace norm of the counter-diabatic driving term $H_\text{\tiny STA}$, which is linked to the energetic cost [cf.~\eq{eq.energeticcost}], in the vicinity of $t\simeq\tau/2$, where the overall dynamics is shown in the inset. Panels \textbf{(b)} and \textbf{(c)} show the dynamics of the position density distribution of the system under the open dynamics at 1\,K temperature, respectively for the quantum protocols 1 and 2. 
	Panel \textbf{(d)} shows the coherence evolution for the quantum protocol 1 (continuous blue line) and 2 (dashed red line) compared to the classical protocols 1 (continuous black) and 2 (dashed black).
	Panel \textbf{(e)} showcases the fidelity of the quantum protocols  for the unitary dynamics (continuous black line) and for the open dynamics (continuous and dashed colored lines) at different temperatures. A fidelity smaller than one indicates that the state of the system is different from the target state, and that possibly only part of the system is transferred from one to the other well. This information is corroborated by the transfer percentage [cf.~\eq{eq.transferperc}], which is shown in Panel \textbf{(f)}. }
	\label{fig:quantumunitary}
\end{figure*}

The results of the open dynamics simulations for the two classical protocols are shown in \fig{fig:classicalunitary}.
In \subf{fig:classicalunitary}{a} 
and \subf{fig:classicalunitary}{b}, we compare the instantaneous numerically-solved eigenvalues of $H_0(t)$ defined in \eq{equ:discretedoublewellsystem} for the continuous (colored lines) and discrete (dashed lines) case for the protocol 1 (panel a) and protocol 2 (panel b).
The comparison between the continuous and discrete case show that our simulations are accurate in the low energy sector encompassing the first $15$ energy levels. An exception to this is given by the first classical protocol, where the system energy goes slightly beyond this regime [cf. \subf{fig:classicalunitary}{e}], then the corresponding results are slightly affected by the approximation.
In \subf{fig:classicalunitary}{c} and \subf{fig:classicalunitary}{d}, we show the evolution of the system's position density $|\braket{x|\psi(t)}|^2$ along with the time-varying potential undergoing the simple control at $1$\,K for protocol 1 and protocol 2 respectively. For protocol 1, the deformation of the potential pushes the system to higher energies, which can be seen by the fact that the system is delocalised over the entire potential, i.e. it is not well localised in just one of the two wells. The system is then cooled down during the slowly restoration of the potential, due to interaction with the cold environment. In the end, one can see the system is mainly localised in the left well with a distribution close to a Gaussian form, which concludes the state transfer protocol. For protocol 2 [cf. \subf{fig:classicalunitary}{d}], the potential is less deformed than that in protocol 1, thus the system gets less excited. Consequently, the final state given by  protocol 2 has a higher fidelity to the corresponding thermal state. In the end, one can see the system is mainly localised in the left well and the state transfer is completed.

To quantify the accuracy of the state transfer, we consider the system energy and the transfer percentage. In \subf{fig:classicalunitary}{e}, we show the comparison between the system energies (continuous and dashed lines) and the corresponding thermal energies (dotted lines) serving as references. The lines in blue and red correspond to protocols operating respectively at 1\,K and 10\,K. The continuous lines represent the protocol 1, while the dashed ones identify protocol 2.
One can see that the final energy is close to the thermal one when protocol 2 is operating at $1$\,K. This is not the case for that protocol 1 when operating at $1$\,K, neither it is for both protocols at $10$\,K, at which temperature the thermal state energy is just below the tip of the barrier, and thus the cold bath fails to localise the system in the finite time $\tau\omega=300$. On the other hand, one can see that the protocol 2 operates better than the protocol 1 at both temperatures: the system energies grow less and are closer to the corresponding thermal ones.
In \subf{fig:classicalunitary}{f}, we report the behaviour of the transfer percentage, which is defined as
\begin{equation}\label{eq.transferperc}
    P(x\leq0)=\int_{-\infty}^0 \D x\,\braket{x|\rho|x},
\end{equation}
for protocol 1 and 2 at 1\,K and 10\,K (we use the same color and dashing as in \subf{fig:classicalunitary}{e}).
At the end of the protocols when the potential is restored to its original shape, the transfer percentage is around 80\% for protocol 1 (90\% for protocol 2) at $1$\,K, while it is  below 60\% for both protocols at $10$\,K.

\subsection{Quantum protocol}

The simulations for the quantum protocols are performed with the CD driving introduced in Eq.~\eqref{equ:STAfullH}, where we set $\delta=0.001$ and $\tau\omega=10$. We underline the difference of the latter timescale with that considered for the classical protocols in Sec.~\ref{sec.classicalprotocol}, which was $\tau\omega=300$. We consider the environmental action as given by the master equation in Eq.~\eqref{equ:systemmasterequation} with the values of the parameters being the same as those considered for the classical protocol. 

\begin{table*}[t]
\caption{Comparison between the state transfer protocols here considered and the corresponding protocol grading under the action of a $1\,$K and 10\,K environment. In particular, we consider the energy cost weighted by Brues angle $\Lambda_\tau^\text{\tiny hs}/ \sin^2\left[\mathcal{L}(\rho_i,\rho_f)\right]$, the time-scale $\omega\tau$, the irreversible entropy production $\Sigma_\text{ir}$, the transfer percentage $P(x\leq0)$, the speed cost $g_\text{\tiny S}$, the quality cost $g_\text{\tiny Q}$, the thermodynamic cost $g_\text{\tiny T}$. Finally, we show the protocol grading $\mQF$.
\label{tab.classical}}
\centering
\begin{tabularx}{0.7\textwidth}{ |c| *{8}{Y|} }
\cline{2-9}
        \multicolumn{1}{c|}{}& \multicolumn{2}{c|}{Classical 1} & \multicolumn{2}{c|}{Classical 2} & \multicolumn{2}{c|}{Quantum 1} & \multicolumn{2}{c|}{Quantum 2} \\
    \hline
     $T$ & $1\,$K & $10\,$K & $1\,$K & $10\,$K & $1\,$K & $10\,$K & $1\,$K & $10\,$K \\
    \hline
    $\omega\tau$ & \multicolumn{2}{c|}{300} & \multicolumn{2}{c|}{300} & \multicolumn{2}{c|}{10} & \multicolumn{2}{c|}{10} \\
    $\Lambda_\tau^\text{\tiny hs}/ \sin^2\left[\mathcal{L}(\rho_i,\rho_f)\right]$ & 0.13 & 0.08 & 0.20 & 0.07 & 2.20 & 1.72 & 2.06 & 1.33 \\
    $\Sigma_\text{ir}$ & 1.37 & 2.36 & 2.23 & 4.33 &0.10 &0.47 & 0.10 &0.46 \\
    $P(x\leq0)$ & 82.39\% & 57.80\% & 91.54\% & 56.32\% &99.98\% &96.10\% &95.45\% &68.28\% \\
    \hline
    $g_\text{\tiny S}$ & 0.88 & 0.90 & 0.86 & 0.90 & 0.90 & 0.91 & 0.90 & 0.92\\
    $g_\text{\tiny Q}$ & 0.27 &  0.09 & 0.36 & 0.10 &0.94 &0.59 &0.89 &0.40 \\
    $g_\text{\tiny T}$ &  0.25 &  0.09 & 0.11 & 0.01 & 0.90 & 0.63 & 0.90 & 0.63 \\
    \hline
    $\mQF$ & 0.06 & 0.007 & 0.03 & 0.0009 & 0.76 &0.34 &0.72 & 0.23 \\
    \hline
\end{tabularx}
\end{table*}

The simulation results are shown in \fig{fig:quantumunitary}. In \subf{fig:quantumunitary}{a}, we compare 
the energetic costs of the CD term, which are computed 
 for the quantum protocols \cite{Campbell_2017,Zheng_2016} with 
\begin{equation}\label{eq.energeticcost}
\norm{H_\text{\tiny STA}(t)}=\sqrt{\tr{H_\text{\tiny STA}^\dag(t) H_\text{\tiny STA}(t)}}.
\end{equation}
One can observe that the majority of the cost using $\alpha_1^\text{\tiny Q}(t)$ (blue line) appears around $t=\tau/2$, i.e. when the potential is symmetric. At this time, the CD term enlarges the energy gaps between eigenstates of the system, thus allowing the system to travel along the trajectories with a high speed and without jumping from one trajectory to the other \cite{Campbell_2017}. Conversely, when employing $\alpha_2^\text{\tiny Q}(t)$ (red line),
the requirement of $\dot{\alpha}_2^\text{\tiny Q}(\tau/2)=0$ imposes a reduction of the instantaneous cost of CD term. 
The evolution of the position density of the system undergoing to the unitary dynamics with $\alpha_1^\text{\tiny Q}$ and $\alpha_2^\text{\tiny Q}$ are shown respectively in \subf{fig:quantumunitary}{b} and in \subf{fig:quantumunitary}{c}, along with the time varying potential with the quantum protocol. As one can see, the system is initially well localised in the right well. Then, at $t=\tau/2$, the state goes in a 
superposition of left and right states, and the position density delocalises over both wells. Eventually, the system is fully transferred to the left well, where the system localises again. 
In \subf{fig:quantumunitary}{d}, we compare the dynamics of the coherence $C_{l_1}$, which is quantified with a $l_1$ measure in the basis of $x'$ \cite{Baumgratz_2104} as  $C_{l_1}(\rho) = \sum_{j\neq k}|\rho_{jk}|$, for the unitary dynamics described by the two quantum protocols and compare them with the two classical ones.
The interaction with the environment reduces the performances of the protocol in two manners. 
First, as it is shown in \subf{fig:quantumunitary}{e}, the fidelity between the final state $\rho_f$  and the target state $\rho_\text{\tiny TG}$ changes
as a decreasing function of the temperature.
Second, the environment inhibits the state transfer form one to the other well. In \subf{fig:quantumunitary}{f}, we show the evolution of the transfer percentage [cf.~\eq{eq.transferperc}], which is the amount of population of the total state that has been transferred from the right to the left well. One can notice that the transfer percentage has a behaviour being similar to that of the fidelity. Both the quantum protocols perform in a similar way, with negligible differences, under unitary evolution when it comes to the fidelity and transfer percentage. 
When comparing the transfer percentage with that  of the classical protocols shown in \subf{fig:classicalunitary}{f}, one clearly see that  quantum protocols are more effective and faster (indeed for the quantum protocol $\tau\omega=10$, while for the classical protocol one has $\tau\omega=300$). Moreover, \subf{fig:quantumunitary}{e} and \subf{fig:quantumunitary}{f} show that the quantum protocol 1 performs, in terms of fidelity and state transfer, better than protocol 2.

\section{Quantification of the protocol}\label{sec:quantification}

Having introduced the classical and quantum protocols for the state transfer, we now 
quantify their performances employing our protocol grading $\mQF$ defined in \eq{equ:quantifier}. For simplicity, in the following, we only consider the environment at the temperature of $1$\,K. Nevertheless, we  report the relevant quantities to compute $\mQF$ also for the case of 10\,K in Table \ref{tab.classical}, and compare them with those for 1\,K.

\begin{figure*}[hbt]
    \stackunder[5pt]{\textbf{(a)}}{\includegraphics[width=0.32\textwidth]{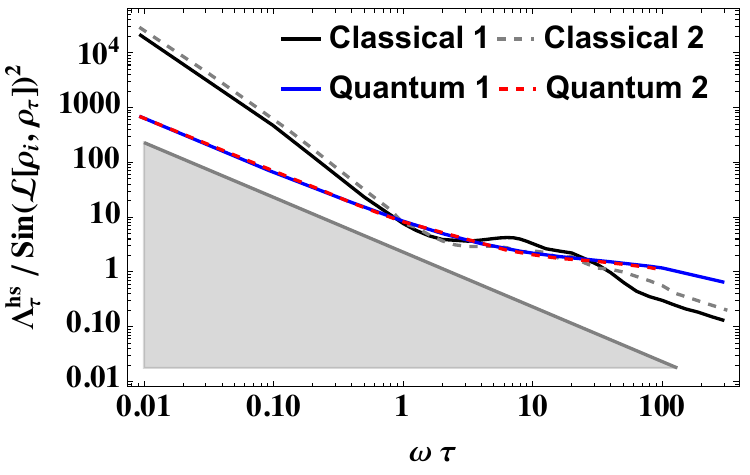}}
    \hfill
    \stackunder[5pt]{\textbf{(b)}}{\includegraphics[width=0.32\linewidth]{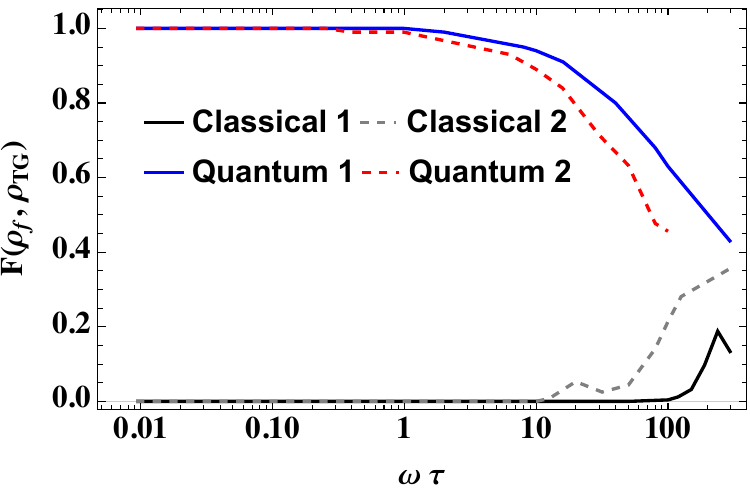}}
    \hfill
    \stackunder[5pt]{\textbf{(c)}}{\includegraphics[width=0.32\linewidth]{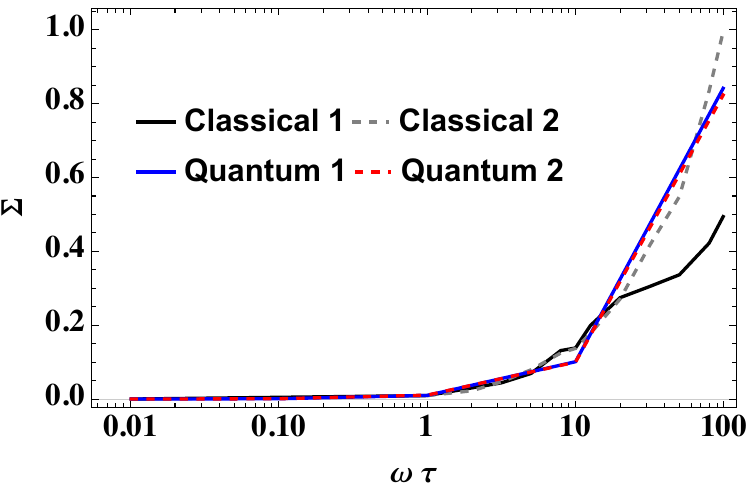}}
	\caption{Quantification of the state transfer protocols for the two classical and two quantum protocols for the temperature of 1\,K. Panel \textbf{(a)} shows the relation  between the inverse of the QSL ($\Lambda_\tau^\text{\tiny hs} / \sin^2\left[\mathcal{L}(\rho_i,\rho_f)\right]$) and the processing time $\omega\tau$. The gray-shaded region corresponds to the values of $\braket{E}_\tau$ and $\tau$ that violate the QSL. Panel \textbf{(b)} shows the achieved fidelity $F(\rho_\tau,\rho_\text{\tiny TG})$ for $\omega\tau=10$ for the quantum protocols and $\omega\tau=300$ for the classical ones. In Panel \textbf{(c)}, we show the evolution of the irreversible entropy production rate $\Pi$. }
	\label{fig:energy-fidelity-entropy}
\end{figure*}

{In \subf{fig:energy-fidelity-entropy}{a}, we show the energetic cost weighted by the Brues angle $\Lambda_\tau^\text{\tiny hs}/ \sin^2\left[\mathcal{L}(\rho_i,\rho_f)\right]$, which is defined in \eq{equ:genericmtqsl}, against the time-scale $\omega\tau$ of different protocols. The gray area identifies the region forbidden by the QSL. Its boundary is characterised by the value of $g_\text{\tiny S}=1$ [cf.~\eq{eq.def.gs}], which corresponds to $\Lambda_\tau^\text{\tiny hs}/ \sin^2\left[\mathcal{L}(\rho_i,\rho_f)\right]=1/\omega\tau$.
The closer the line of the protocol is to the gray region, the better $g_\text{\tiny S}$.  
In this figure, we clearly see the advantages of quantum protocols against the classical ones for small processing times $\omega\tau<1$. This is due to the fact that classical protocols fail to produce distinguishable states (\ie~$\sin^2\left[\mathcal{L}(\rho_i,\rho_f)\right]\sim 0$), while the quantum protocols are always able  (\ie~$\sin^2\left[\mathcal{L}(\rho_i,\rho_f)\right]\sim 1$). 
On the other hand, for large processing times ($\omega\tau>40$), although both kind of protocols almost always produce distinguishable states, the environmental effects become no longer negligible. 
Instead, we see that quantum advantages disappear and the classical protocols behave better the quantum ones. 
Finally, for intermediate processing times ($1<\omega\tau<40$), the intertwined behaviours of lines fail to provide any useful information on their performances.}

{In \subf{fig:energy-fidelity-entropy}{b}, we show the fidelity $F(\rho_\tau,\rho_\text{\tiny TG})$ between the final state $\rho_\tau$  to the target state $\rho_\text{\tiny TG}$ [cf.~\eq{equ:targetstate}]  against the processing time $\omega\tau$.  As it is expected, for short processing times, the quantum protocols can achieve the task perfectly, \ie~transferring the state while preserving the correct information, while the classical protocols fail completely. For long processing times, both quantum and classical protocols result in similar fidelity due to thermalisation. One can notice that quantum protocol 1 performs better than quantum protocol 2 (where we imposed that $\dot H_0(\tau/2)=0$); while classical protocol 2 performs better than classical protocol 1 (since there is less tilting involved). }

{In \subf{fig:energy-fidelity-entropy}{c}, we show  the irreversible entropy production $\Sigma_\text{ir}$ [cf.~\eq{equ:ratetoproduction}] against $\omega\tau$. 
It indicates how far the system is driven away from the reversible dynamics and it quantifies the thermodynamic cost $g_\text{\tiny T}$.
We see that the effects of 
 the environment becomes prominent when 
 $\omega\tau>1$.}

\begin{figure}[t]
  \centering
\includegraphics[width=0.9\linewidth]{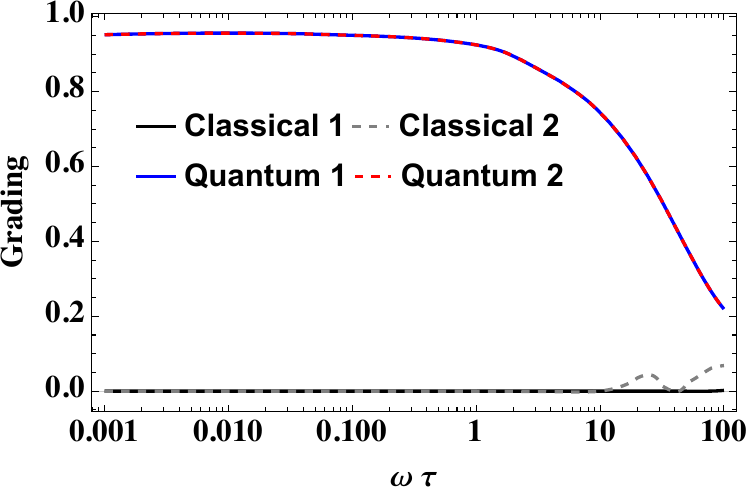}
	\caption{ 
	The protocol grading $\mQF$ for the four considered state transfer protocols under the action of the environment at 1\,K at varying protocol time-scale.}
\label{fig:protocolgrading}
\end{figure}

{Having computed all terms in the protocol grading $\mQF$ given by \eq{equ:quantifier}, we show in \fig{fig:protocolgrading} the values of $\mQF$ for both the quantum and the classical protocols. For $\omega\tau<1$, we see a strong difference between the classical ($\mQF\sim 0$) and the quantum ($\mQF\sim 1$) protocols, which narrows for larger values of $\omega\tau$. This is due to the fact that quantum protocols well perform for time smaller that the decoherence time ($\tau_\text{dec}\sim\hbar^2/\gamma m k_\text{\tiny B} T \Delta_x^2\sim0.7$, where we took $\Delta_x$ as the distance between the two minima of double-well potential); while the classical protocols perform better with a more adiabatic time-scale ($\omega\tau\gg1$). }

\section{Conclusions}\label{sec:conclusions}
In this paper, we design the protocol grading $\mQF$ that depends on a finite set of fundamental physical quantities, and it is used as a figure of merit of the performances of a process, by taking into consideration the speed, the fidelity and the thermodynamic cost of the transfer process.
We compute the protocol grading for four different state transfer protocols in a double-well potential. These transfer a quantum superposition from one well to the other. Such a process can be successfully performed by employing a quantum protocol with the help of the counter-diabatic driving, while with classical processes the information about the initial superposition is washed away. Furthermore, quantum processes allow a state transfer that is  quicker and more accurate, which is reflected by  higher value for the protocol grading parameter.

\acknowledgements
We thank Steve Campbell for many useful discussions. We acknowledge financial support from the H2020-FETOPEN-2018-2020 project TEQ (grant nr. 766900), the DfE-SFI Investigator Programme (grant 15/IA/2864), the Royal Society Wolfson Research Fellowship (RSWF\textbackslash R3\textbackslash183013), the Leverhulme Trust Research Project Grant (grant nr.~RGP-2018-266), the UK EPSRC (grant nr.~EP/T028106/1). M.A.C. acknowledges support from the Lise Meitner Programme (number M2915).

\bibliography{Summary}.bib

\appendix

\section{Analogue with Landau-Zener Model}\label{apd:landauzenermodel}

The simplest analogy to our proposed model for the quantum protocol of the state transfer is given by the Landau-Zener model \cite{Landau:1932uk,zener}. The model considers a two-level system with an Hamiltonian reading
\begin{equation}
  H_\text{\tiny LZ} = \Delta \sigma_z + g(t)\sigma_x,
\end{equation}
where $\Delta$ determines the minimal energy gap between the two energy levels, and $g(t)$ is a linear control function. In \subf{fig:LZmodel}{a}, we plot the eigenvalues of $H_\text{\tiny LZ}$  in terms of $g$ (continuous lines). As one can see, if we prepare initially the system in its ground state ($\ket{-}$) and then drive the system non-adiabatically by changing $g$, the system jumps from the ground state to the excited one ($\ket{+}$). This is most likely to happen  when $g$ is around $0$, which corresponds to the minimum energy gap. 

To circumvent the issue, we can introduce a counter-diabatic term to the original Hamiltonian \cite{Chen_2010,Campbell_2017}. The new Hamiltonian now reads
\begin{equation}
  H^\text{new}_\text{\tiny LZ} = H_\text{\tiny LZ} + H_\text{\tiny STA},
\end{equation}
where 
\begin{equation}
  H_\text{\tiny STA} = -\frac{\dot g(t) \Delta}{2 (\Delta^2 + g(t)^2)}\sigma_y.
\end{equation}
With such a counter-diabatic term, the ground state trajectory becomes the finite-time solution of the new Hamiltonian $H^\text{new}_\text{\tiny LZ}$.
This is due to the increased energy gap imposed by $H_\text{\tiny STA}$, as it is indicated by the arrows in \subf{fig:LZmodel}{b}.
The increased energy gap allows the system to remain in the ground state without jumping to the excited state as we change $g$. We underline that following the ground state (blue continuous line) one switch the state from $\ket{-}$ to $\ket{+}$ (dashed lines in \subf{fig:LZmodel}{a}). This is essentially the state transfer we want to simulate in our model. By changing the value of $g$ in time, the population of the state $\ket{-}$ moves to the state $\ket{+}$. This is pictured in \subf{fig:LZmodel}{c} where the dimensions of the colored disks represent the amount of population of the two states (blue for $\ket{-}$ and red for  $\ket{+}$).

However, the dynamics looks differently if we consider the position of the system rather than the spin. Our framework imposes the relation between the position operator and the spin operator $J_x$ based on the HP transformation. Taking this Landau-Zener model as example, we can take $\sigma_x$ as a close analogue of the position operator, and monitor the population that changes in ``positions'', i.e. from $-1$ to $+1$ (corresponding to the eigenstates $\ket{-}$ and $\ket{+}$). If we prepare our initial state as $\ket{-}$, we can see the state transfers from one position to the other position, as shown in \subf{fig:LZmodel}{c}.

\begin{figure}[tb]
  \stackunder[5pt]{\textbf{(a)}}{\includegraphics[width=0.49\linewidth]{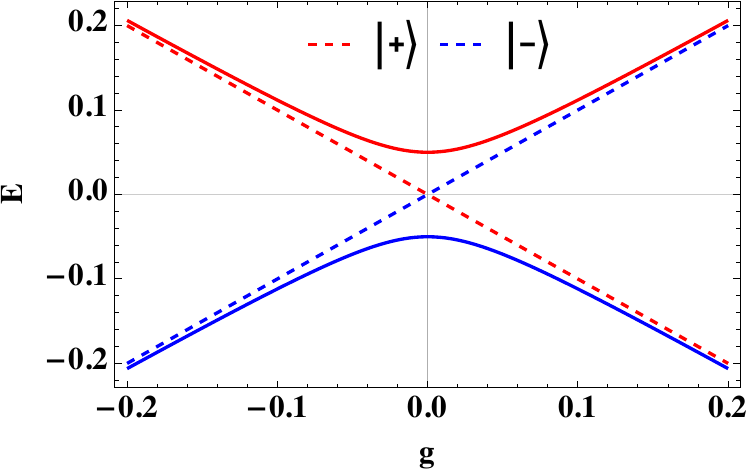}}
\stackunder[5pt]{\textbf{(b)}}{\includegraphics[width=0.49\linewidth]{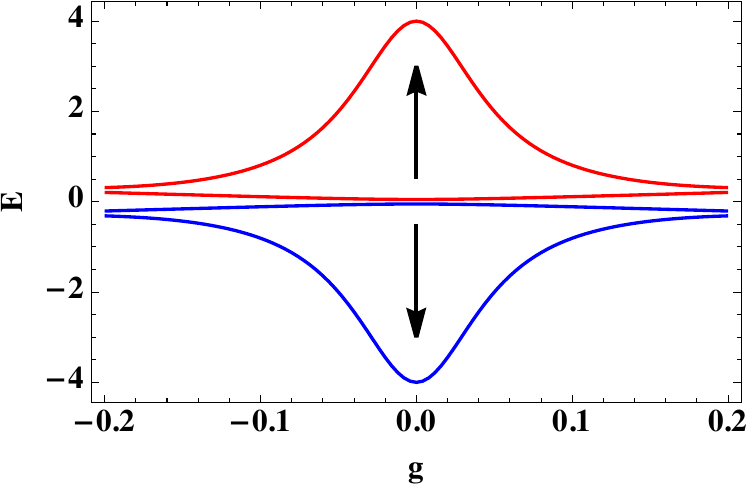}}\\\vspace{1em}
\stackunder[5pt]{\textbf{(c)}}{\includegraphics[width=0.6\linewidth]{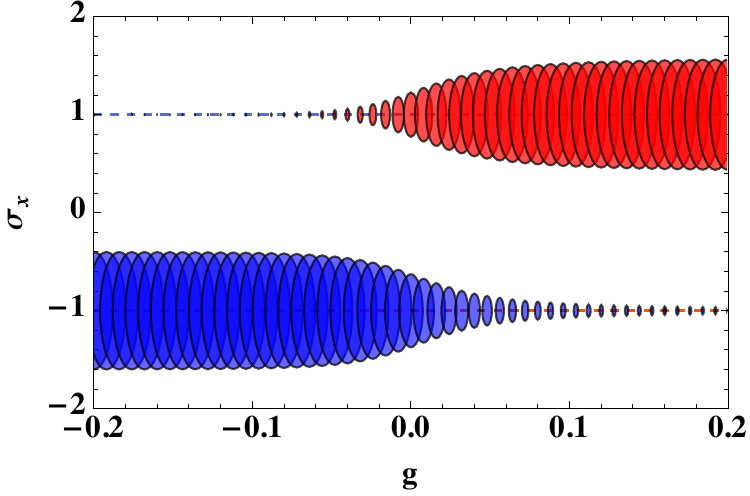}}
  \caption{Illustration of the state transfer with Landau-Zener model. Here, Panel \textbf{(a)} shows the change of energy against the tuning parameter $g$. Panel \textbf{(b)} shows the change of energy for states $\ket{+}$ and $\ket{-}$ as the STA is implemented. Panel \textbf{(c)} shows how population changes in $\sigma_x$ basis, if we prepare $\ket{-}$ as the initial state. Here we used $\Delta = 0.05$.}
  \label{fig:LZmodel}
\end{figure}

\section{Decompose the dissipator into reversible and irreversible parts}\label{apd:decomposerates}

In this appendix, we compute, along the lines of \cite{Santos_2018}, the contributions to the entropy rates corresponding to the  following term of the Caldeira-Leggett dissipator in \eq{equ:cldissipator}
\begin{equation}
  D[\rho] = [x,\{p,\rho\}] = [x,f(\rho)],
\end{equation}
where $
  f[\rho] = f^\dag[\rho] = \{p,\rho\}$.
As $ D[\rho]$ contains the operator $x$, we express the dissipator in the Husimi-Q representation. In particular, the Husimi-Q function is defined in \eq{husimi.def}
and we have the following correspondences \cite{Gardiner_2004}
\begin{equation}
  \begin{aligned} 
    C_x[\rho]=[x,\rho]~\leftrightarrow&~C_x(Q)=\frac{1}{\sqrt{2}}(\partial_{\alpha^\ast} -\partial_\alpha)Q, \\
    C_p[\rho]=[p,\rho]~\leftrightarrow&~C_p(Q)=-\frac{i}{\sqrt{2}}(\partial_{\alpha^\ast} +\partial_\alpha)Q, \\
    f[\rho]=\{p,\rho\}~\leftrightarrow&~f(Q) =\frac{i}{\sqrt{2}}(2\alpha^\ast - 2\alpha -\partial_{\alpha^\ast} +\partial_\alpha)Q, 
  \end{aligned}
\end{equation}
where we define two currents $C_x(Q)$ and $C_p(Q)$. 
One can work out the reverse correspondences,
\begin{equation}  
  \begin{aligned}
    \partial_{\alpha}Q &= -\frac{1}{\sqrt{2}}(C_x(Q)-i C_p(Q)), \\
    \partial_{\alpha^\ast}Q &= \frac{1}{\sqrt{2}}(C_x(Q)+i C_p(Q)), \\
  \end{aligned}
\end{equation}
and thus rewrite $f(Q)$ in terms of currents,
\begin{equation}\label{B4}
  f(Q) = i\sqrt{2}(\alpha^\ast - \alpha)Q - i C_x(Q).
\end{equation}
With these correspondences, we can have the dissipator in phase space,
\begin{equation}\label{equ:dissipatorqfunc}
  D[\rho]~\leftrightarrow~ D(Q) = C_x(f(Q)).
\end{equation}
Given the definition of the Wehrl entropy 
 in \eq{equ:wehrlentropy}, we can rewrite its rate component corresponding to the term $D[\rho]$ as
\begin{equation}
    \frac{\D S}{\D t} = \int \D\alpha \int \D\alpha^\ast\,\frac{1}{Q} f(Q) C_x(Q).
\end{equation}
Employing \eq{B4}, we get
\begin{equation}
\frac{\D S}{\D t} = i\int \D\alpha \int \D\alpha^\ast\, \underbrace{\sqrt{2}(\alpha^\ast - \alpha) C_x(Q)}_{\text{flux}} + \underbrace{\frac{1}{Q} \abs{C_x(Q)}^2}_{\text{irreversible}},
\end{equation}
where we use the correspondences, integrate by parts and take $C_x(Q) = -C_x(Q)^\ast$.
By working backwards, we find the corresponding decomposition in density matrix representation, which is given by
\begin{equation}
    D[\rho] = -i[x,[x,\rho]] - i\sqrt{2}[x,\rho a-a^\dag\rho],
\end{equation}
being \eq{equ:decomposerates} of the main text.

Now, the irreversible entropy production rate $\Pi$ is associated with an even function of the current, and the entropy flux rate $\Phi$ is associated with an odd function of the current \cite{Landi_2021}.
According to this argument, we can separate the Wehrl entropy rate into 
\begin{equation}
  \begin{aligned}
    \Pi&=i\int \D\alpha \int\D\alpha^\ast\, \frac{1}{Q} \abs{C_x(Q)}^2, \\
    \Phi&=i\int \D\alpha \int\D\alpha^\ast\, \sqrt{2}(\alpha - \alpha^\ast) C_x(Q),
  \end{aligned}
\end{equation}
which are respectively the irreversible entropy production rate and entropy flux rate.

\end{document}